\documentclass{aa}
\usepackage{graphicx}
\usepackage[varg]{txfonts}
\usepackage{natbib}
\usepackage{soul}
\newlength{\linwx}
\usepackage{color}

\begin{document}

\title{Influence of sub- and super-solar metallicities on the compositions of solid planetary building blocks}

\author{
Bertram Bitsch \inst{1},
\and
Chiara Battistini \inst{2}
}
\offprints{B. Bitsch,\\ \email{bitsch@mpia.de}}
\institute{
Max-Planck-Institut f\"ur Astronomie, K\"onigstuhl 17, 69117 Heidelberg, Germany
\and
Zentrum f\"ur Astronomie der Universit\"at Heidelberg, Landessternwarte K\"onigstuhl, K\"onigstuhl 12, 69117 Heidelberg, Germany
}
\abstract{The composition of the protoplanetary disc is thought to be linked to the composition of the host star, where a higher overall metallicity of the host star provides more building blocks for planets. However, most of the planet formation simulations only link the stellar iron abundance [Fe/H] to planet formation and the iron abundance in itself is used as a proxy to scale all elements. On the other hand, large surveys of stellar abundances show that this is not true. We use here stellar abundances from the GALAH surveys to determine the average detailed abundances of Fe, Si, Mg, O, and C for a broad range of host star metallicities with [Fe/H] spanning from -0.4 to +0.4. Using an equilibrium chemical model that features the most important rock forming molecules as well as volatile contributions of H$_2$O, CO$_2$, CH$_4$ and CO, we calculate the chemical composition of solid planetary building blocks around stars with different metallicities. Solid building blocks that are formed entirely interior to the water ice line (T>150K) only show an increase in Mg$_2$SiO$_4$ and a decrease in MgSiO$_3$ for increasing host star metallicity, related to the increase of Mg/Si for higher [Fe/H]. Solid planetary building blocks forming exterior to the water ice line (T<150K), on the other hand, show dramatic changes in their composition. In particular the water ice content decreases from around $\sim$50\% at [Fe/H]=-0.4 to $\sim$6\% at [Fe/H]=0.4 in our chemical model. This is mainly caused by the increasing C/O ratio with increasing [Fe/H], which binds most of the oxygen in gaseous CO and CO$_2$, resulting in a small water ice fraction. Planet formation simulations coupled with the chemical model confirm these results by showing that the water ice content of super-Earths decreases with increasing host star metallicity due to the increased C/O ratio. This decrease of the water ice fraction has important consequences for planet formation, planetary composition and the eventual habitability of planetary systems formed around these high metallicity stars.
}
\authorrunning{Bitsch \& Battistini}\titlerunning{Influence of sub- and super-solar metallicities on planetary compositions}\maketitle

\section{Introduction}
\label{sec:Introduction}

Observations of exoplanets have reveled that close-in super-Earths are the most abundant type of planet within 1 AU \citep{2011arXiv1109.2497M, 2018AJ....156...24M, 2018arXiv180502660Z}. These super-Earths have typically masses of a few Earth masses and planets within the same system could be of similar size \citep{2018AJ....155...48W} and thus, presumably mass. Even though recent analysis have questioned this trend \citep{2019arXiv190702074Z}. 

These super-Earths exist around many different type of stars, even around small M-dwarfs \citep{2016Natur.533..221G,2017Natur.542..456G}. In addition, there seems to be no trend with host star metallicity regarding the occurrence rate of these super-Earths \citep{2012Natur.486..375B}. This seems to make super-Earths the most robust outcome of planet formation in nature. 

However, \citet{2019MNRAS.485.3981S} found that planets below 30 Earth masses seem to have a positive correlation in the host star metallicity-period-mass diagram, meaning that the mass of the planet increases with both, host star metallicity and orbital period. The trend that planetary masses increase with host star metallicity is not surprising, as giant planets are more common around host stars with higher metallicity \citep{2004A&A...415.1153S, 2005ApJ...622.1102F, J2010}. The trend with orbital distance could also be explained by host star metallicity. If more building blocks are available, planetary embryos at larger distances can also grow faster and form planets by the accretion of pebbles \citep{2015A&A...582A.112B}, which could explain these trends and could also give rise to the giant planet eccentricity distribution \citep{2018arXiv180206794B}. 

If the planetary radius through transit observations and the planetary masses through RV detections are known, the mean density of the planet can be calculated (e.g. see \citealt{2018ApJ...865...20D} for the Trappist-1 system). This can give important information about the planetary composition through interior structure models \citep{2007ApJ...665.1413V, 2007Icar..191..337S, 2007ApJ...669.1279S, 2007ApJ...659.1661F, 2007Icar..191..453S, 2008ApJ...673.1160A, 2013PASP..125..227Z, 2016AJ....152..160B}, which can give important information about the formation history of the planet and their evolution \citep{2018A&A...610L...1V}. Planets with mean densities close to the terrestrial planets most likely formed in the inner regions of the protoplanetary disc without significant accretion of water ice, while planets with lower densities could harbour a significant fraction of water ice \citep{2019arXiv190604253Z}. Of course, the planetary radius measurements could be greatly influenced by planetary atmospheres.

The formation pathways of these super-Earth, on the other hand, is still under debate and a lot of different theories have been proposed \citep{2007ApJ...654.1110T, 2008ApJ...685..584I, 2009ApJ...699..824O, 2010ApJ...719..810I, 2010MNRAS.401.1691M, 2012ApJ...751..158H, 2014A&A...569A..56C,2014ApJ...780...53C, 2015A&A...578A..36O, 2016ApJ...817...90L, 2017MNRAS.470.1750I}. Studies that form super-Earths in the outer parts of the disc and then let them migrate into the inner regions have the advantage that they can naturally explain water rich super-Earths \citep{2018MNRAS.479L..81R, 2019arXiv190208772I, 2019A&A...624A.109B, 2019arXiv190600669S}. However, the formation pathway of forming super-Earths in the outer disc and then migrating them inwards depends crucially on the water fraction within the protoplanetary disc, because it influences the growth and migration pathway of planets \citep{2016A&A...590A.101B}. 

Planet formation simulations that include chemical models have mostly used solar like composition. These models either operate in the classical planetesimal accretion scenario \citep{2010ApJ...715.1050B, 2014A&A...570A..36M, 2015A&A...574A.138T, 2015BAAA...57..251R} or with planet formation assisted at planet migration traps \citep{2017MNRAS.464..428A, 2019A&A...627A.127C} or include pebble accretion \citep{2017MNRAS.467.2845A, 2017MNRAS.469.4102M}. Some of these models additionally focus on the chemical abundances of the planetary atmospheres \citep{2017MNRAS.467.2845A, 2017MNRAS.469.4102M, 2019A&A...627A.127C}, while others mainly focus on the chemical composition of rocky/icy super-Earths \citep{2017MNRAS.464..428A}. Here we want to focus on the chemical composition of solid planetary building blocks, which is set by the underlying chemical composition of the disc and not focus on the underlying physics needed for planet formation theories.

The process of planet formation is thought to happen within the protoplanetary disc surrounding the newly formed star. A key assumption in models of planet formation is that the elemental abundances of the protoplanetary disc are directly correlated to the host star, meaning that the disc and the host star have the same elemental composition. However, in planet formation simulations it is not very often taken into account that stellar elemental abundances show different trends when compared to [Fe/H]. In addition most studies of planet formation only focused on solar like composition for the different elements.

These different elemental trends in stars are caused by the fact that different elements are produced in different kind of stars at different times in the galactic evolution \citep{1957RvMP...29..547B}. Elements like O, Mg, Si, Ca, Ti (also known as $\alpha$-elements) as well as S, are mostly produced in massive stars during their evolution and then released in the interstellar space during supernovae explosions (Type II supernovae, SNII): this means that this enrichment is expected to happen on a short timescale \citep{1986A&A...154..279M}. Low mass stars instead can pollute the interstellar medium on a longer time scale via, for example, Type Ia supernovae (SNIa), where significant quantities of Fe are released and only small fraction of $\alpha$-elements (e.g. \citealt{2002Ap&SS.281...25T}). SNIa are created from white dwarfs in binary systems, resulting in a time delay relative to SNII (e.g. \citealt{2009A&A...501..531M}) in a range between 0.3 to 1 Gyr (e.g. \citealt{2009NewA...14..638V}). It is important to know that even if different elements share the same production site (like SNII), the difference in mass of the progenitor star and/or the explosion parameters matters \citep{2006ApJ...653.1145K, 2011ApJ...729...16K} for the elemental yields and thus how they are incorporated to molecular clouds that form new stars.

Several works showed the trend of $\rm \alpha$-elements in respect to [Fe/H] in different sample sizes and different parts of the Galaxy (e.g. \citealt{2012A&A...545A..32A, 2014A&A...570A.122S, 2014A&A...562A..71B, 2017A&A...605A..89B, 2018MNRAS.478.4513B}). Elements in different elemental groups (light elements, iron-peak elements and neutron-capture elements) might have different trends simply because they are formed in different sites (see for example \citealt{2014A&A...562A..71B, 2015A&A...577A...9B, 2016A&A...586A..49B}). However these trends have only been taken into account in a very small sample of studies related to planet formation and composition. 

Recent studies have focused on the mineralogy of super-Earth planets and how it could vary with stellar abundance \citep{2017ApJ...845...61U, 2018ApJ...853...83H, 2019arXiv190705506P} or by building the planet very close to its host stars where the temperatures are very high changing the molecules that form the planet \citep{2019MNRAS.484..712D}. These studies took only the formation interior to the water ice line (T>150K) into account and \citet{2019arXiv190705506P} focused on the mantle mineralogy. \citet{2019arXiv190910058L} linked detailed stellar observations of HD219134 to constrain the properties of its planetary companions and determine the core to mantle ratio of these planets. The work by \citet{2017A&A...608A..94S} followed a different approach by studying the chemical composition of planetary building blocks around thin and thick disc stars in general, showing that the planetary building blocks differ in their iron and water mass fraction. \citet{2019A&A...622A..49C} used stellar population synthesis simulations to predict the chemical composition of planetary building blocks using the same chemical model as \citet{2017A&A...608A..94S}, finding similar results.

Here we combine detailed stellar abundance measurements of C, O, Mg, Si and S\footnote{S is not directly measured in the GALAH survey, but we assume it scales with [Fe/H] the same as Si, as shown from observations in \citet{2002A&A...390..225C}}, from the GALAH survey to calculate the bulk composition of solid planetary building blocks. We do not distinguish in this study planetary host stars and stars that have no detected planets, because the formation of planets seems to be an universal process, where on average nearly all stars should host planets within 1 AU \citep{2018AJ....156...24M}. For simplification, we mainly focus only on the individual solid planetary building blocks rather than incorporating them in a complex planet formation model (see appendix~\ref{ap:formation}). We thus focus on solids that either form completely interior or exterior to the water ice line.

Our paper is structured as follows. In section~\ref{sec:methods} we describe our chemical model and the calculation of the solid planetary building blocks. In section~\ref{sec:placomp} we show the compositions of solids formed around host stars with different metallicities and for solids formed completely interior and exterior to the water ice line. In section~\ref{sec:formation} we apply the chemical model used to calculate the solid planetary building blocks to a planet formation model and show the final water ice content of the formed planets. We then discuss the implications of our results in section~\ref{sec:disc} and summarize our findings in section~\ref{sec:summary}.

\section{Methods}
\label{sec:methods}

\subsection{Stellar abundances}

For our study we use stellar abundances derived for the second data release of the GALAH (Galactic Archeology with HERMES) survey \citep{2018MNRAS.478.4513B} that contains 342,682 stars. GALAH is a high-resolution (R $\sim$ 28,000), large-scale stellar spectroscopic survey of Milky Way stars in the magnitude range 12 < V < 14 probing mostly the Galactic thin and thick disk but also a consistent number of halo stars. The Hermes spectrograph observes in four discrete optical wavelength bands (4713-4903\AA, 5648-5873\AA, 6478-6737\AA and 7585-7887\AA) meaning that for our studies sulfur is not included in the observations (the triplet S I lines commonly used for chemical abundance determination are located around 6756 \AA).

Stellar parameters and elemental abundances are derived using the data-driven approach of $The \ Cannon$ \citep{2015ApJ...808...16N} that first uses a training sample to derive several stellar labels that are going to be propagated to the entire sample (see \citealt{2018MNRAS.478.4513B} for the details). The training sample is analysed using SME ({\it Spectroscopy Made Easy}, \citealt{1996A&AS..118..595V, 2017A&A...597A..16P}) that performs spectrum synthesis with 1D LTE (Local Thermal Equilibrium) hydrostatic stellar atmosphere MARCS models. It is important to mention that GALAH DR2 incorporates non-LTE line formation for several elements using departure coefficients from LTE. In respect to our study, O \citep{2016MNRAS.455.3735A}, Mg \citep{2015A&A...579A..53O}, Si \citep{2017MNRAS.464..264A} and Fe \citep{2016MNRAS.463.1518A} are corrected for 1D non-LTE.

We selected only dwarf stars ($T_{\rm eff}$ > 5000 and log $g$ > 4.0) that have a good parameter flag value from {\it The Cannon} (\texttt{flag\_cannon == 0}). At this point we select for each element only the stars for which {\it The Cannon} gives a good abundance flag (\texttt{flag\_x\_fe == 0}): this lead to a sample of 142,000 stars with good O abundance, 27,638 stars for C, 148,765 for Mg and 121,105 for Si. 

The fact that so few stars from the original sample have good C abundances, is caused by the combination of detection limits and the flagging algorithm of {\it The Cannon}. In fact only one C line is used for the abundance determination (at 6588 \AA) and only a small training sample for C is available (see Tab.2 in \citealt{2018MNRAS.478.4513B}).

\begin{figure}
 \centering
 \includegraphics[scale=0.5]{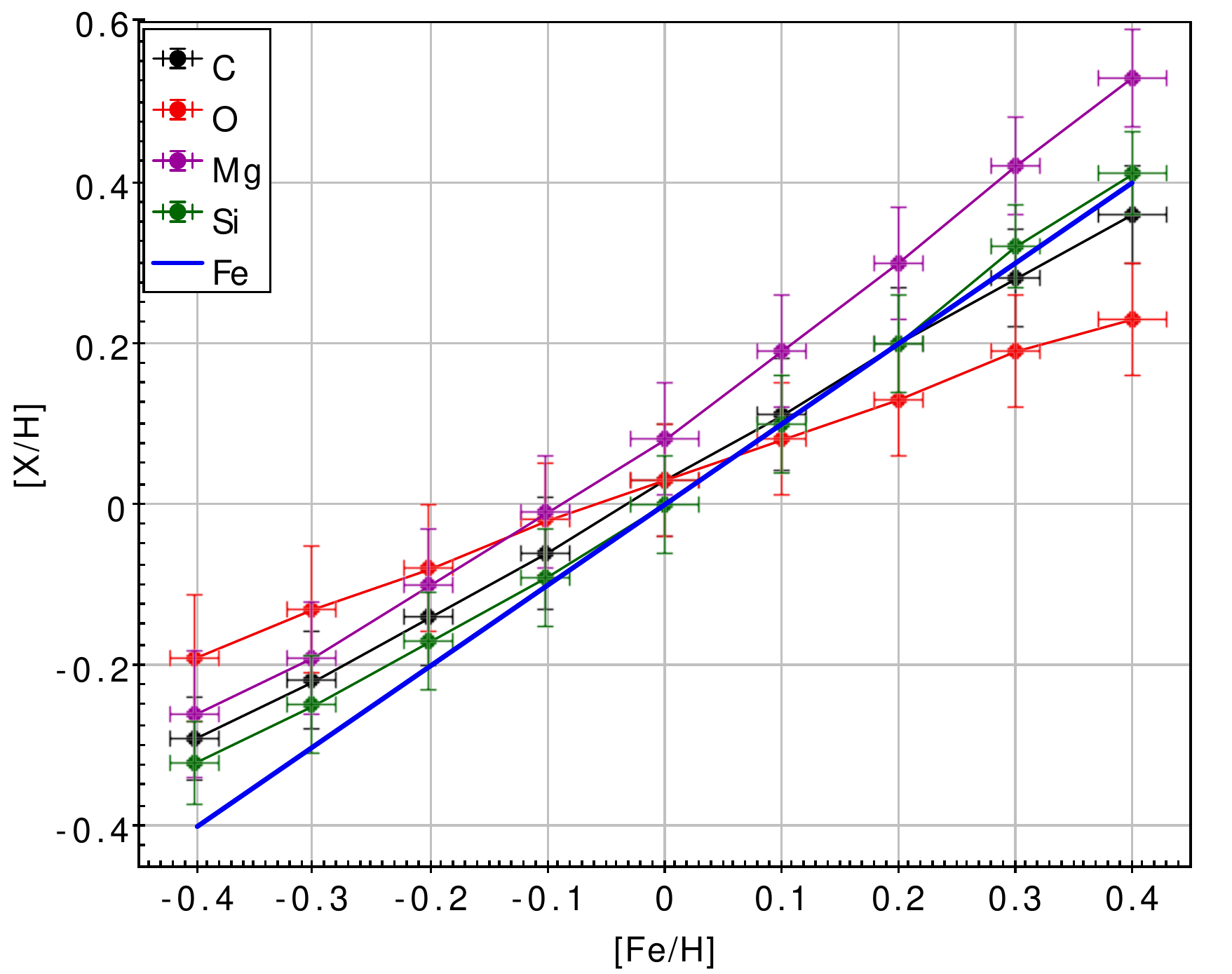}
 \caption{Stellar abundances of Mg, Si, O and C as function of [Fe/H] for the stars in our sample. The stellar abundances have been observed in the GALAH survey \citep{2018MNRAS.478.4513B}. The error bars are the mean deviations of the observations. The data is given in tab.~\ref{tab:abu}. Sulfur (not shown) scales in the same way as silicon \citep{2002A&A...390..225C}.
   \label{fig:abundances}
   }
\end{figure}

We then grouped the stars selected for each element in metallicity bins of 0.1 dex as shown in Fig.~\ref{fig:abundances}, determining the mean value and dispersion of [X/H] in each bin, where X is C, O, Mg and Si. It is clear from Fig.~\ref{fig:abundances} that the assumption of scaling all the [X/H] (with X one of the element in this study) as [Fe/H] is not correct, especially for stars in the more metal-poor and metal-rich regime of our metallicity range. The trends of [X/H] vs. [Fe/H] for the stars selected in this study are presented in the Figure~\ref{fig:stars} in appendix~\ref{sec:abu}.

Even if our values are corrected for non-LTE, it is important to remember that they are based on 1D models. Recently new determination of non-LTE values are derived using 3D model atmospheres. In this respect \citet{2019A&A...622L...4A} shows new abundances for C, O and Fe corrected for 3D non-LTE for 187 F and G stars in the Galactic disk and halo. It seems that even if the correction for O are more severe in the metal-rich regime, the correction from our 1D non-LTE values should not be big, because our sample is made of dwarf stars where this effect is not that large.

As previously said, there are no S abundance determination from GALAH and we assume that it scales as Si as shown in \citet{2002A&A...390..225C}. The same trend is present in several studies \citep{2005A&A...441..533C, 2011A&A...530A.144J, 2016PASJ...68...81T, 2017A&A...604A.128D} regarding dwarfs and giants, reassuring us to scale S with the Si abundance trends.

\subsection{Chemical model}

In order to account for the chemical composition of the planet, we include only the major rock and ice forming species. The mixing ratios (by number) of the different species as a function of the elemental number ratios is denoted X/H and corresponds to the abundance of element X compared to hydrogen for solar abundances, which we take from \citet{2009ARA&A..47..481A} and are given as follows: He/H = 0.085; C/H = $2.7\times 10^{-4}$; O/H = $4.9\times 10^{-4}$; Mg/H = $4.0\times 10^{-5}$; Si/H = $3.2\times 10^{-5}$; S/H = $1.3\times 10^{-5}$; Fe/H = $3.2\times 10^{-5}$.

These different elements can combine to different molecular species. We list these species, as well as their condensation temperature and their volume mixing ratios $v_{\rm Y}$ in Table~\ref{tab:species}. More details on the chemical model can be found in \citet{2017MNRAS.469.4102M} and \citet{2018MNRAS.479.3690B}.

To account for different host star metallicities, we scale the solar values from \citet{2009ARA&A..47..481A} with the corresponding scaling from Fig.~\ref{fig:abundances}. In addition, sulfur scales the same as silicon \citep{2002A&A...390..225C}.

As we focus on solid formation close to the water ice line, the solids in our simulation still form at temperatures larger than 70K, meaning that CO, CH$_4$ and CO$_2$ are in gaseous form and they thus do not contribute to the solid planetary building blocks. However, including the gaseous CO and CO$_2$ binds a lot of oxygen, which is then not available to form water ice, in contrast to the study by \citet{2017A&A...608A..94S} and \citet{2019A&A...622A..49C}. This binding of oxygen can greatly reduce the amount of available water around stars with high C/O ratios. We discuss how the amount of carbon bound in molecules that do not bind any oxygen (e.g. methane) influences our results in appendix~\ref{ap:carbon}.

\begin{table*}
\centering
\begin{tabular}{c|c|c}
\hline
Species (Y) & $T_{\text{cond}}$ {[}K{]} & $v_{\text{Y}}$ \\ \hline \hline
CO & 20  & 0.45 $\times$ C/H  \\[5pt]
CH$_4$ & 30 & 0.45 $\times$ C/H   \\[5pt]
CO$_2$ & 70 & 0.1 $\times$ C/H  \\[5pt]
H$_2$O & 150 & O/H - (3 $\times$ MgSiO$_3$/H + 4 $\times$ Mg$_2$SiO$_4$/H + CO/H \\
& & + 2 $\times$ CO$_2$/H + 3 $\times$ Fe$_2$O$_3$/H + 4 $\times$ Fe$_3$O$_4$/H) \\[5pt]
Fe$_3$O$_4$ & 371 & (1/6) $\times$ (Fe/H - S/H) \\[5pt]
FeS & 704 & S/H \\[5pt]
Mg$_2$SiO$_4$ & 1354 & Mg/H - Si/H  \\ [5pt]
Fe$_2$O$_3$ & 1357 & 0.25 $\times$ (Fe/H - S/H) \\ [5pt]
MgSiO$_3$ & 1500 & Mg/H - 2 $\times$ (Mg/H - Si/H)  \\  \hline
\end{tabular}
\caption[Condensation temperatures]{Condensation temperatures and volume mixing ratios of the chemical species. Condensation temperatures for molecules are taken from \citet{2003ApJ...591.1220L}. For Fe$_2$O$_3$ the condensation temperature for pure iron is adopted \citep{2003ApJ...591.1220L}. Volume mixing ratios $v_{\rm Y}$ (i.e. by number) are adopted for the species as a function of disc elemental abundances (see e.g. \citealt{2014ApJ...791L...9M}). We note that the Mg abundance is always larger than the Si abundance.}
\label{tab:species}
\end{table*}

\subsection{Planet formation}

As outlined in the introduction, the formation of super-Earths can happen through many different formation channels. However, all core accretion theories agree that solid material, either in the form of pebbles or planetesimals (for a review see \citealt{Johansen2017}), must be accreted to reach masses of a few Earth masses. In this work, we focus on the elemental and molecular composition of planetary building blocks formed completely interior (T>150K) and completely exterior (T<150K)to the water ice line, before applying it to a planet formation model. In principle growing planets migrate through the disc (for a review see \citealt{2013arXiv1312.4293B}), but accretion, especially in the pebble accretion scenario, can easily be fast enough that the planet is fully grown before it starts to migrate significantly \citep{2015A&A...582A.112B}. However, in reality, planets formed exterior to the water ice line can also cross the water ice line during their assembly, which allows them to have a lower water content compared to planets formed completely exterior to the water ice line \citep{2019A&A...624A.109B}, which can help to explain the formation of the Trappist-1 system \citep{2019arXiv190600669S}. The planetary composition in this case is just a mixture between the composition acquired exterior and interior to the water ice line. We show in section~\ref{sec:formation} and appendix~\ref{ap:formation} the results of simple planet formation model including pebble accretion, disc evolution and planet migration. The final water ice content of planets in this model is directly correlated to the temperature at which the solids are accreted and is as such always a mixture between the minimum and maximal allowed water ice content of the model, which we discuss in section~\ref{sec:placomp}.

\section{Composition of solid planetary building blocks formed completely interior or exterior to the water ice line}
\label{sec:placomp}

In this section we show the elemental and molecular composition of solid planetary building blocks formed entirely interior or exterior to the water ice line (but interior to the CO$_2$ ice line) to give an overview of the extreme cases. We then apply the same chemical model in section~\ref{sec:formation} and study the compositions of planets that grow by pebble accretion and migrate through the disc, where they can cross the ice lines. We first discuss the elemental and molecular composition of solids formed at [Fe/H]=0 and then move to lower and higher [Fe/H] values. We then summarize the findings in Fig.~\ref{fig:Moleculetrend} and Fig.~\ref{fig:Elementtrend}.

\subsection{Solar metallicity}

\begin{figure}
 \centering
 \includegraphics[scale=0.87]{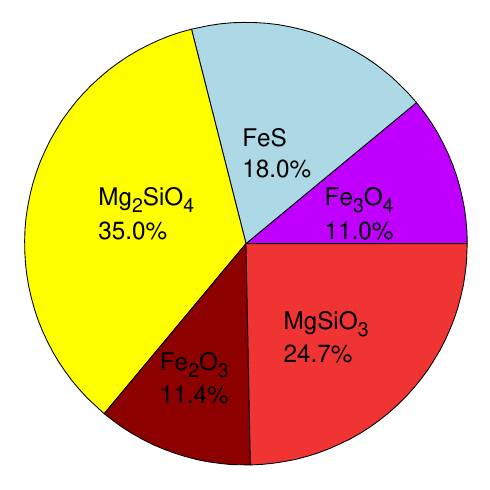}
 \includegraphics[scale=0.87]{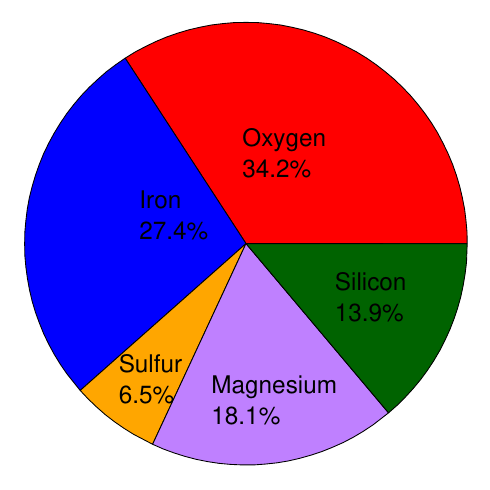}
 \caption{Molecular (left) and elemental mass fractions (right) of solid planetary building blocks formed completely interior to the water ice line (T>150K) around a star with [Fe/H]=0. The mass of the building blocks is dominated by Mg$_2$SiO$_4$ and MgSiO$_3$, while the elemental masses are dominated by oxygen and iron, making up around 60\% of the total mass of the planet.
    \label{fig:solarT200}
   }
\end{figure}

In Fig.~\ref{fig:solarT200} we show the molecular and elemental mass fractions of solid planetary building blocks formed entirely interior to the water ice line (T>150K) at [Fe/H]=0.0. The mass of the solid planetary building blocks on a molecular level is dominated by Mg$_2$SiO$_4$ and MgSiO$_3$, while the elemental masses are dominated by oxygen and iron.

The total elemental masses from our chemical model are similar to the Earth's abundance \citep{McDonough95}, however, our model shows a larger magnesium fraction compared to the Earth. This is related to the usage of the magnesium abundance from the Galactic abundances at [Fe/H]=0, which is larger than the solar values. This results in an overabundance of magnesium for our planetary composition. If we instead use directly the solar abundances \citep{2009ARA&A..47..481A} our model is much closer to the Earth's composition, but does not match to exactly. We attribute these differences to our simple chemical model which only traces the main carriers of material. In addition, our formation model does not account for the impact history on Earth, where impactors originated from different regions in the solar system.

\begin{figure}
 \centering
 \includegraphics[scale=0.87]{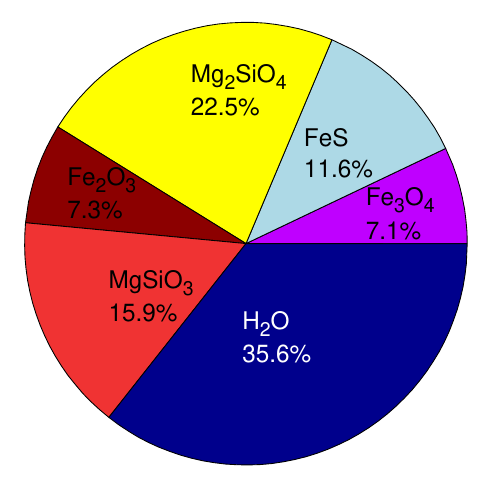}
 \includegraphics[scale=0.87]{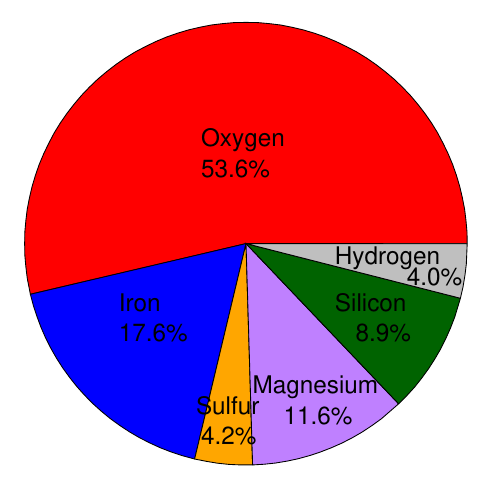}
 \caption{Molecular (left) and elemental mass fractions (right) of solid planetary building blocks formed completely exterior to the water ice line (T<150K) around a star with [Fe/H]=0. The molecular level is dominated by H$_2$O and Mg$_2$SiO$_4$, while oxygen and iron clearly dominate the elemental composition, making up nearly 70\% of the planetary mass.
    \label{fig:solarT100}
   }
\end{figure}

In Fig.~\ref{fig:solarT100} we show the molecular and elemental mass fractions of solid planetary building blocks that formed entirely exterior to the water ice line (T<150K). The solid mass is now mainly dominated by water ice (H$_2$O) and thus oxygen. The total elemental composition is then obviously not Earth like at all, confirming that the Earths did not form exterior to the water ice line, even though the water ice line can evolve interior to 1 AU during the evolution of the protoplanetary disc \citep{2011ApJ...738..141O, 2015A&A...575A..28B}. In the pebble accretion scenario this problem could be solved by the growth of Jupiter that blocks the inflow of water rich particles into the inner disc \citep{2016Icar..267..368M}, even though the water ice line sweeps interior to 1 AU after 1 Myr according to viscous disc evolution models \citep{2011ApJ...738..141O, 2015A&A...575A..28B, 2015arXiv150303352B}.

\subsection{Sub-solar metallicities}

\begin{figure}
 \centering
 \includegraphics[scale=0.87]{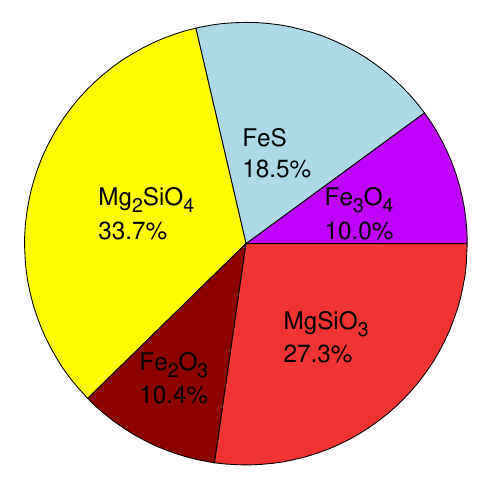}
 \includegraphics[scale=0.87]{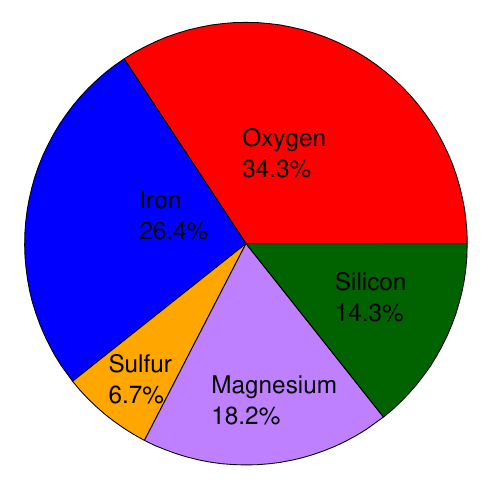}
 \includegraphics[scale=0.87]{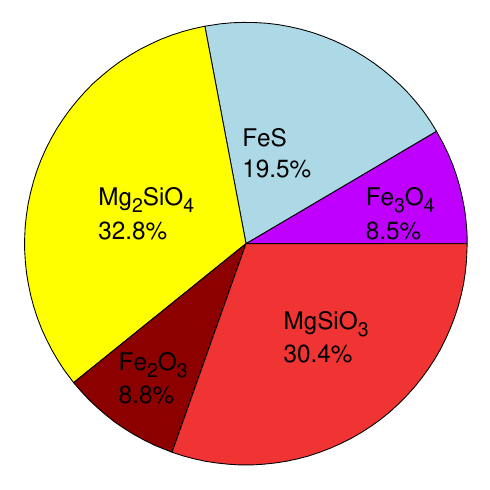}
 \includegraphics[scale=0.87]{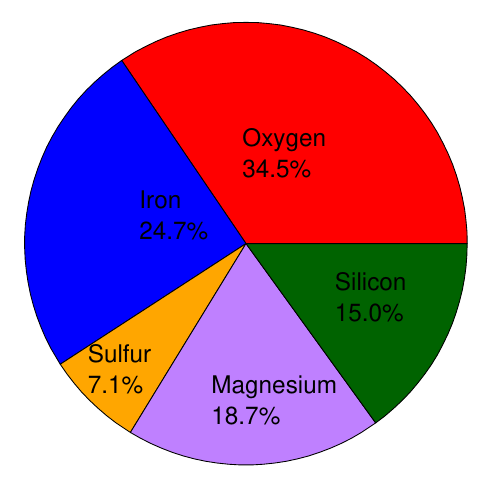}
 \caption{Molecular (left) and elemental mass fractions (right) of solid material formed completely interior to the water ice line (T>150K) around a star with [Fe/H]=-0.2 (top) and [Fe/H]=-0.4 (bottom). As for [Fe/H]=0 (Fig.~\ref{fig:solarT200}), the molecular composition is dominated by Mg$_2$SiO$_4$ and MgSiO$_3$, while the elemental composition is dominated by oxygen and iron, making up around 60\% of the total mass of the planet.
    \label{fig:subsolarT200}
   }
\end{figure}

In Fig.~\ref{fig:subsolarT200} we show the molecular and elemental mass fractions of solid planetary building blocks formed entirely interior to the water ice line (T>150K) around stars with sub-solar [Fe/H]. The overall molecular and elemental abundances are very similar to solids formed around stars with [Fe/H]=0. This is caused by very similar slopes in the abundance trends for silicon, magnesium, iron and sulfur (Fig.~\ref{fig:abundances}). Oxygen, on the other hand, is overabundant compared to the other elements, which has no influence on the amount of the major rock forming species (Mg$_2$SiO$_4$, FeS, Fe$_3$O$_4$, Fe$_2$O$_3$ and MgSiO$_3$) in our chemical model. Thus the composition of solids formed interior to the water ice line around stars with sub-solar metallicity is very similar to stars formed around solar metallicity.

\begin{figure}
 \centering
 \includegraphics[scale=0.87]{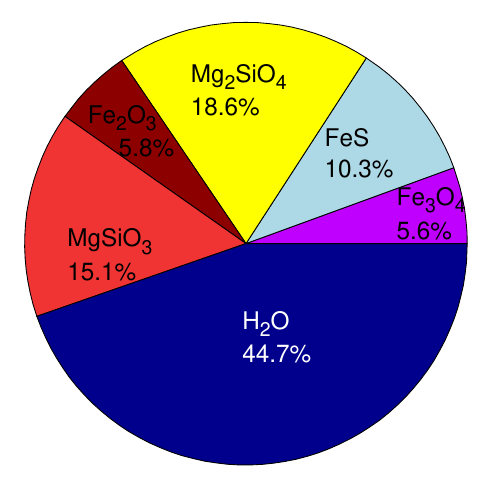}
 \includegraphics[scale=0.87]{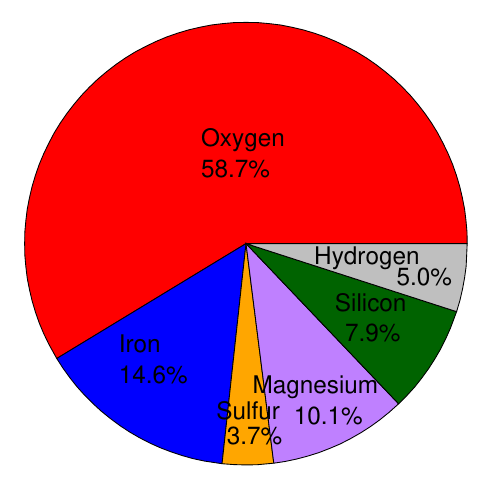}
 \includegraphics[scale=0.87]{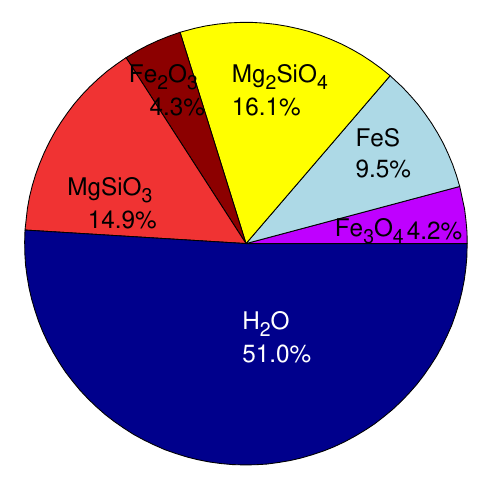}
 \includegraphics[scale=0.87]{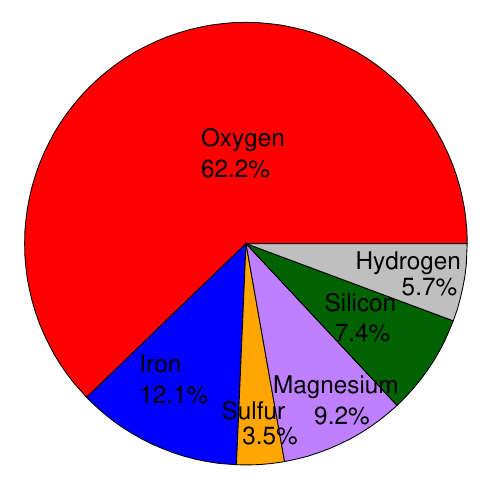}
 \caption{Molecular (left) and elemental mass fractions (right) of solid planetary building blocks formed completely exterior to the water ice line (T<150K) around a star with [Fe/H]=-0.2 (top) and [Fe/H]=-0.4 (bottom). As for [Fe/H]=0 (Fig.~\ref{fig:solarT200}), the molecular composition is dominated by H$_2$O, while the elemental composition is dominated by oxygen and iron, where the iron content decreases for lower overall metallicities. 
    \label{fig:subsolarT100}
   }
\end{figure}

In Fig.~\ref{fig:subsolarT100} we show the molecular and elemental abundances of solid planetary building blocks formed exterior to the water ice line (T<150K) for sub-solar metallicities. For lower and lower [Fe/H], water ice becomes more and more dominant for the total mass budget of the solids and with it, oxygen. This is clearly related to the fact that the oxygen abundances is enhanced compared to the other elements at low [Fe/H], see Fig.~\ref{fig:abundances}. In our chemical model this implies that the relative abundances of rock forming molecules are similar to stars with [Fe/H]=0, but the iron content decreases for lower overall metallicities. This is related to the overabundance of oxygen allowing the formation of more H$_2$O, which thus becomes more and more dominant for lower [Fe/H]. As a consequence the water ice fraction in the solid planetary building blocks can be above 50\% for [Fe/H]=-0.4.

\subsection{Super-solar metallicities}

\begin{figure}
 \centering
 \includegraphics[scale=0.87]{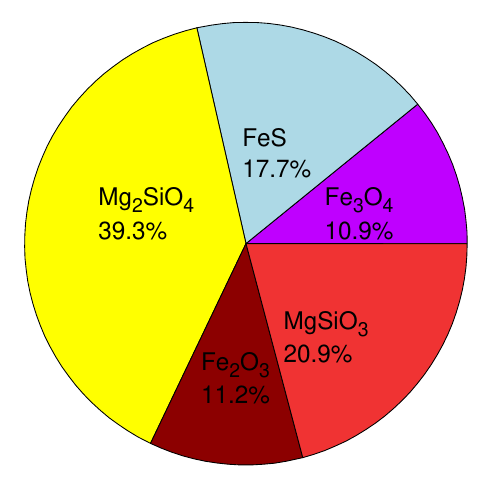}
 \includegraphics[scale=0.87]{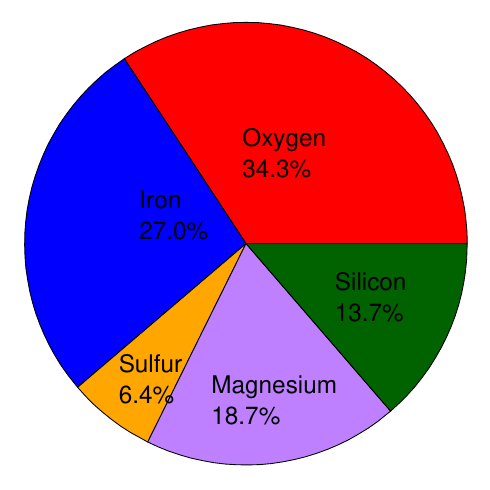}
 \includegraphics[scale=0.87]{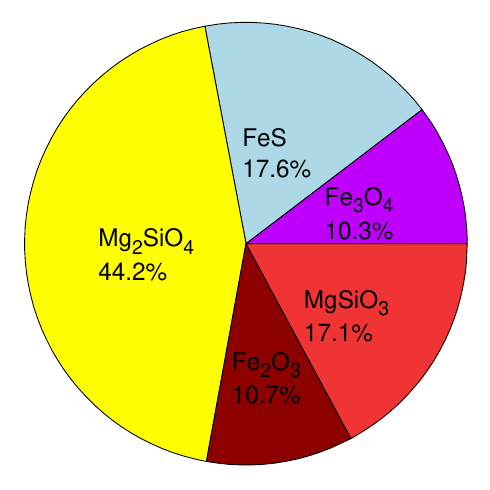}
 \includegraphics[scale=0.87]{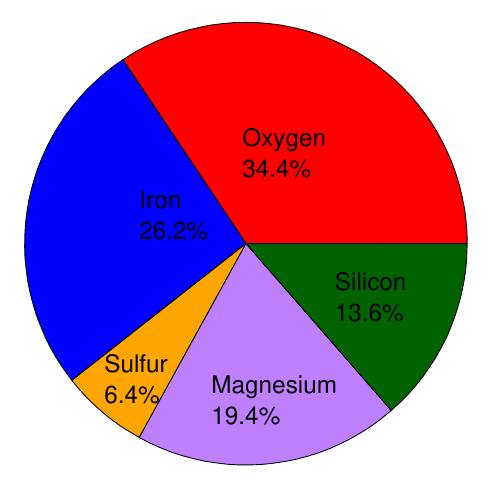}
 \caption{Molecular (left) and elemental mass fractions (right) of solid planetary building blocks formed completely interior to the water ice line (T>150K) around a star with [Fe/H]=+0.2 (top) and [Fe/H]=+0.4 (bottom). As for [Fe/H]=0 (Fig.~\ref{fig:solarT200}), the molecular composition is dominated by Mg$_2$SiO$_4$ and MgSiO$_3$, while the elemental composition is dominated by oxygen and iron, making up around 60\% of the total mass of the planet, as for all [Fe/H] values. 
    \label{fig:supersolarT200}
   }
\end{figure}

In Fig.~\ref{fig:supersolarT200} we display the molecular and elemental mass fractions of solid planetary building blocks formed interior to the water ice line (T>150K) around stars with super-solar metallicity. As can be seen from Fig.~\ref{fig:abundances}, the magnesium abundance increases strongest with [Fe/H], resulting in an overall increase of Mg$_2$SiO$_4$ in the mass fraction of the formed solids. However, the overall mass fractions for the different elements only changes slightly, because the relative mass of Mg and O inside of the rock forming molecules MgSiO$_3$ and Mg$_2$SiO$_4$ changes only slightly as well, leaving the overall elemental trend with [Fe/H] quite constant.

Additionally, the oxygen abundance is reduced compared to the other elements, but there is still enough oxygen to fully oxidize iron, magnesium and silicon. This implies that the reduction of the oxygen abundance relative to the other elements has no influence on the elemental composition of solids formed entirely interior to the water ice line. If the galactic trends for these elements continue for even larger [Fe/H] abundances, we expect a deviation from the here derived mass fractions of the elements inside the solids formed interior to the water ice line, because not enough oxygen to fully oxidize Fe, Mg and Si might be available.

\begin{figure}
 \centering
 \includegraphics[scale=0.87]{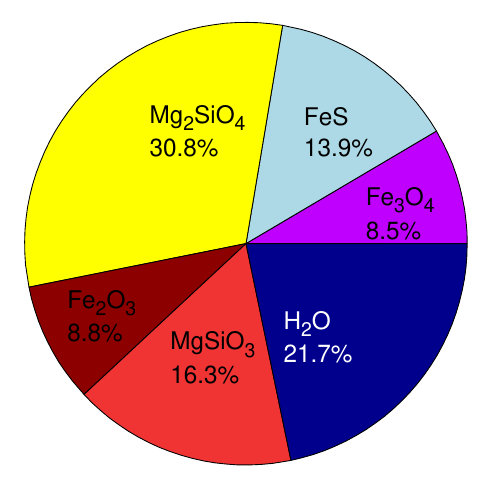}
 \includegraphics[scale=0.87]{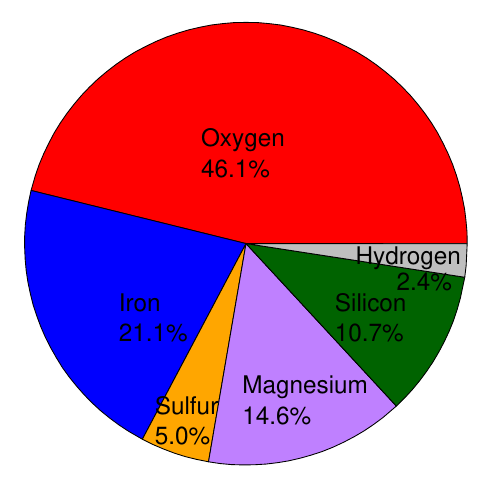}
 \includegraphics[scale=0.87]{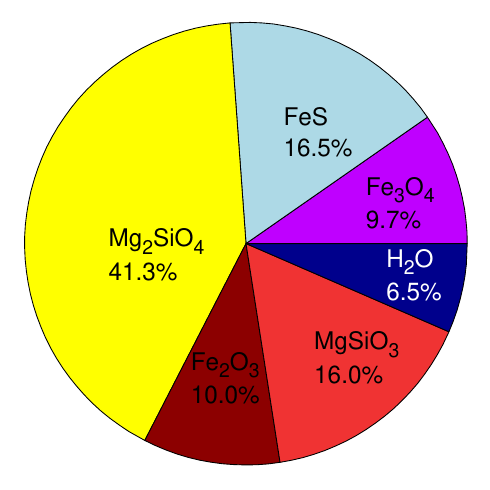}
 \includegraphics[scale=0.87]{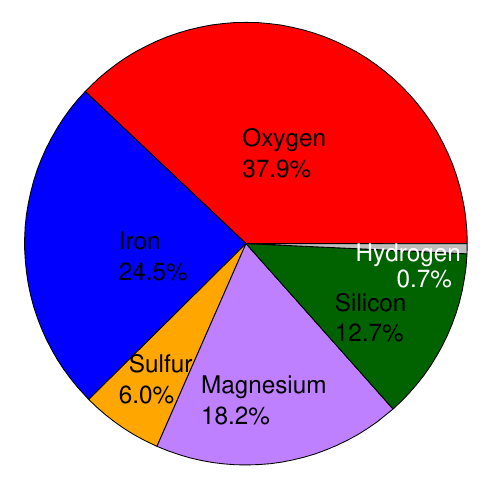}
 \caption{Molecular (left) and elemental mass fractions (right) of solid planetary building blocks formed completely exterior to the water ice line (T<150K) around a star with [Fe/H]=+0.2 (top) and [Fe/H]=+0.4 (bottom). In contrast to the mass fractions of solids formed around lower metallicity stars exterior to the water ice line, the dominant molecular mass is not water ice, but Mg$_2$SiO$_4$. In fact, the water ice content reduces to just a few percent for solids formed around stars with [Fe/H]=0.4. The dominant elemental mass fraction is oxygen and iron, however the oxygen fraction is not as elevated as for planets formed around stars with low [Fe/H], see Fig.~\ref{fig:subsolarT100}.
    \label{fig:supersolarT100}
   }
\end{figure}

In Fig.~\ref{fig:supersolarT100} we show the elemental and molecular mass fractions of solid planetary building blocks formed entirely exterior to the water ice line (T<150K) around stars with super-solar metallicity. In contrast to solids formed around stars with low [Fe/H] exterior to the water ice line (Fig.~\ref{fig:subsolarT100}, the water ice fraction of the solids has decreased dramatically. In fact for [Fe/H]=0.4, the water ice mass fraction is only of a few percent.

The low water ice mass fraction is caused by the reduced oxygen abundances compared to the C, Si, Mg, Fe and S abundances around stars with large [Fe/H], see Fig.~\ref{fig:abundances}. This lower oxygen abundances still allows Si, Mg and Fe to be fully oxidized, binding a lot of oxygen. In addition, the large carbon abundance (Fig.~\ref{fig:CO}) binds a lot of oxygen into CO and CO$_2$, leaving only very little oxygen that can then form water ice (see table~\ref{tab:species}). This results in a very low water ice abundance in solids formed exterior to the water ice line around metal rich stars. As a consequences, the elemental abundances show an increase of Fe, S, Mg, Si and a reduction of H and O with increasing [Fe/H] due to the reduced availability of water ice. If the same trend for all elements holds true for even larger [Fe/H], our chemical model predicts that water ice would not be available at overall higher metallicities than [Fe/H]>0.4.

\section{Planet formation}
\label{sec:formation}

In this section we show the results of a simple planet formation model based on pebble accretion to study the final position, masses and water ice content of formed planets. The details of the planet formation model are given in appendix~\ref{ap:formation} and follow standard models used in the past (e.g. \citealt{2015A&A...582A.112B}). Our model contains a decreasing pebble flux with time, an evolving protoplanetary disc model which cools in time, meaning that the ice lines move inward in time and type-I migration for the growing planetary cores. We start our planetary embryos with 0.01 Earth masses. The growth of the planets stops at the pebble isolation mass, when planets start to open a small gap in the protoplanetary discs, which stops the inward flux of pebbles and thus accretion \citep{2014A&A...572A..35L, 2018arXiv180102341B}. We do not include gas accretion and type-II migration, because our work focuses on the solid composition of material and thus our planet formation model focuses on the formation of super-Earths.

Past simulations have shown that the water ice content in the protoplanetary disc can greatly influence the disc structure and the planet formation pathway regarding growth and migration \citep{2016A&A...590A.101B}. However constructing a detailed disc model based on the different abundances is beyond the scope of this work. We therefore rely on the disc model derived for solar metallicities presented in \citet{2015A&A...575A..28B} and use this for all our simulations. This approach might not be as accurate, but it allows an easy comparison between the different scenarios, because the planet migration rates are the same due to the same disc model.

\subsection{Solar metallicity}

\begin{figure*}
 \centering
 \includegraphics[scale=0.7]{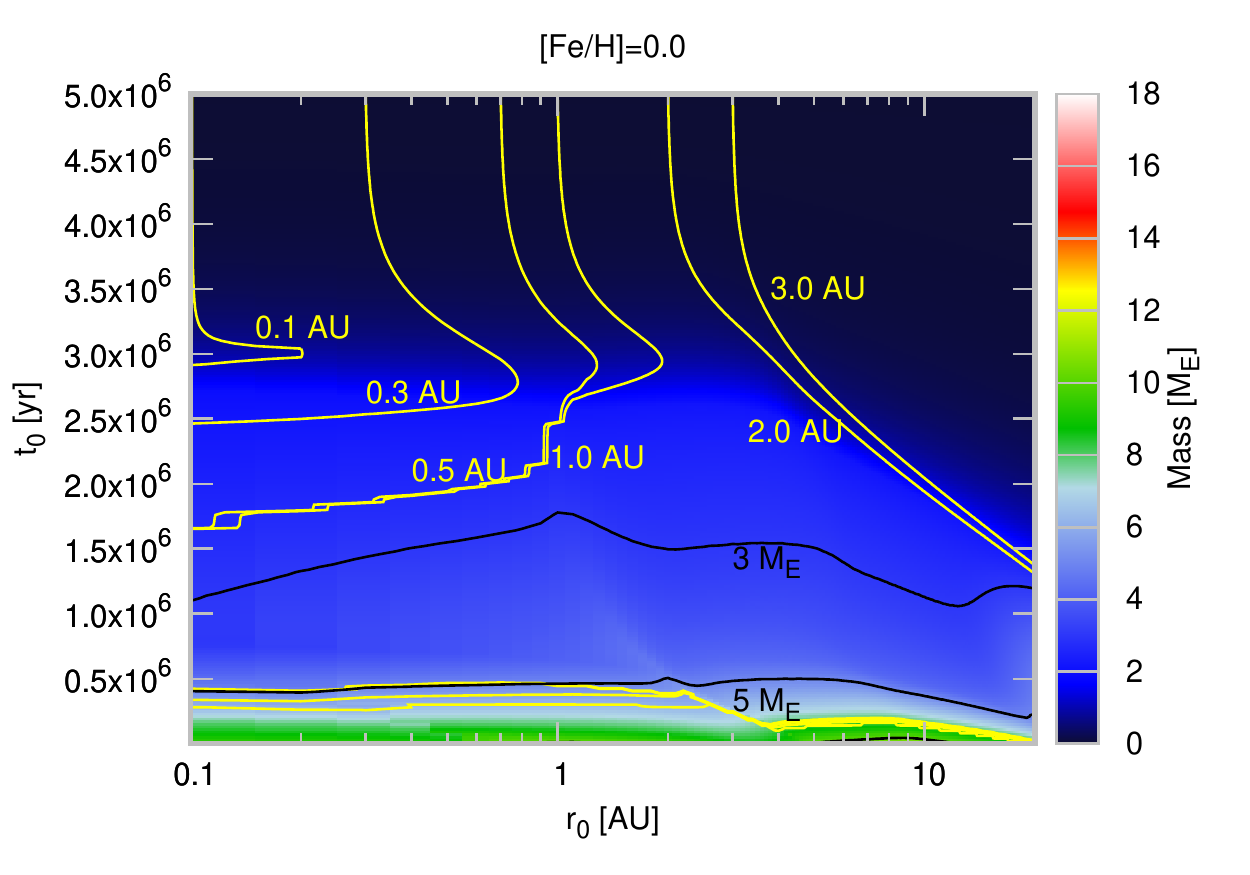} \quad
 \includegraphics[scale=0.7]{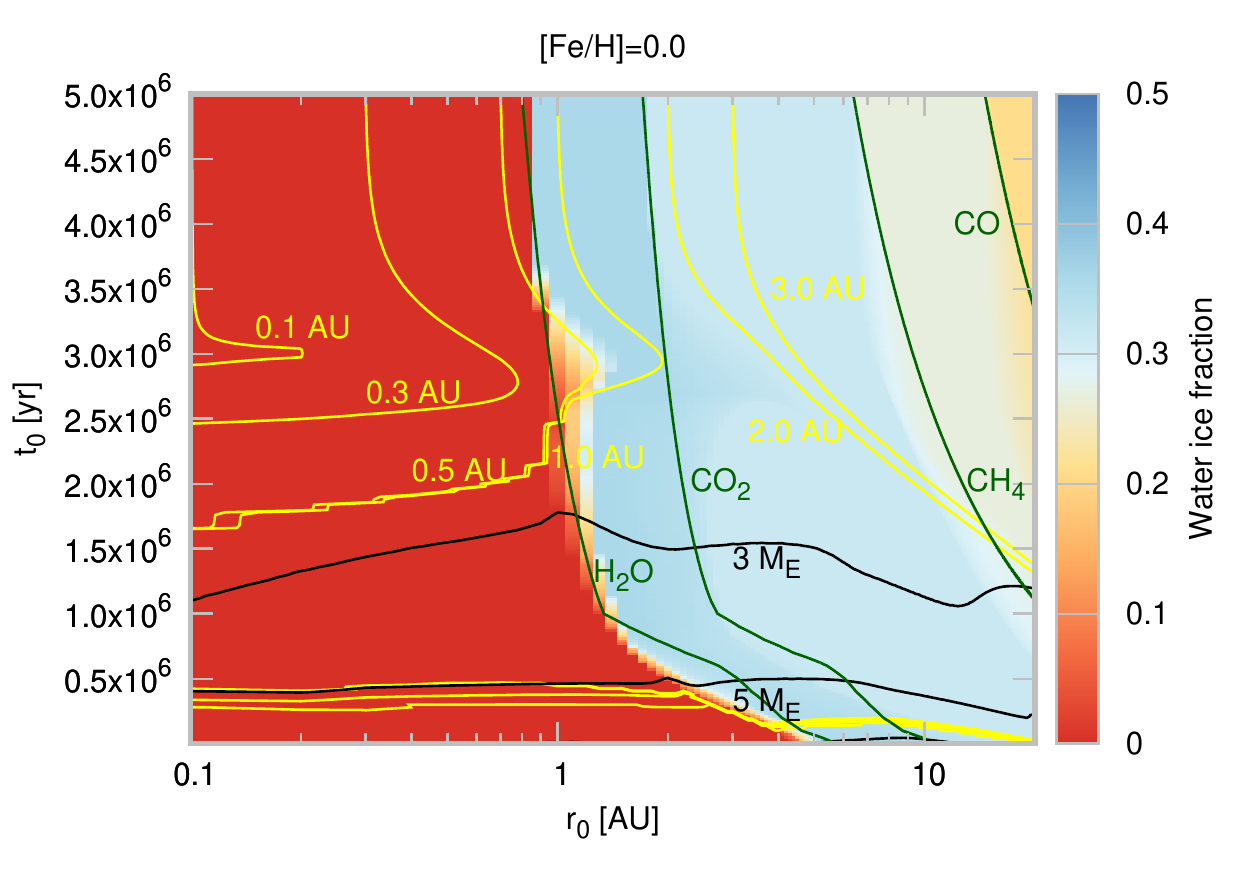}
 \caption{Final planetary masses (left) and final water ice content (right) of the planets formed in our simulations as function of the planetary starting position $r_0$ and starting time in the disc $t_0$ around a star with [Fe/H]=0.0. The yellow lines indicate the final planetary positions in AU after 5 Myr and the black lines mark the final planetary masses in Earth masses. The green lines mark the ice lines of H$_2$O, CO$_2$, CH$_4$ and CO. Planets forming very early in the disc migrate all the way down to 0.1 AU. The majority of planets is trapped in the region of outward migration until the end of the disc's lifetime. The decrease in the water ice fraction at large initial distances is caused by the accretion of CO$_2$, CH$_4$ and CO into the planet, which reduces the relative water ice fraction of the planet, but increase the total ice fraction of the planet.
    \label{fig:PlanetsFeH0}
   }
\end{figure*}

In Fig.~\ref{fig:PlanetsFeH0} we show the final planetary mass (left) and the corresponding water ice content of the planets (right) as function of the starting position $r_0$ and starting time $t_0$ of the planet in the disc around a star with [Fe/H]=0 at the final time of 5 Myr. The yellow and black lines mark the final planetary position and masses, while the green lines mark the ice lines of H$_2$O, CO$_2$, CH$_4$ and CO.

Planets growing in the early stages of the disc (small $t_0$) receive the largest pebble flux and thus grow fastest. In addition, in the early stages of the disc evolution, the pebble isolation mass, at which pebble accretion stops, is highest (e.g. \citealt{2015A&A...582A.112B, 2019A&A...624A.109B}). This results in the most massive planets that can reach masses of up to 5-10 Earth masses within this model ($t_0< 500$ kyr). Due to the model-choice of a constant Stokesnumber for the pebbles, pebbles are much smaller in the outer disc where the gas density is lower, reducing the pebble accretion rate for planets further away from the star.

As the pebble flux decreases in time, the planetary growth rate reduces. Additionally, the disc cools and it's aspect ratio reduces, resulting in a smaller pebble isolation mass at later times. Thus planets formed at large initial times $t_0$ grow slower and to lower masses than planetary embryos injected at earlier starting times.

During their growth, the planets migrate through the disc. The used disc model \citep{2015A&A...575A..28B} features a region of outward migration exterior to the water ice line, where planets can be trapped as they grow. These planets could then only be released from the disc when the disc dissipates and the region of outward migration can only hold small planets \citep{2015A&A...575A..28B}. As a result nearly all of the planets in our simulations are trapped exterior to 0.1 AU, which is the inner edge of our disc model. 

However, planetary systems most likely contain more than one planet, which then results in a different dynamical history of the system. The N-body simulations of \citet{2019arXiv190208772I} and \citet{2019A&A...623A..88B} use the same disc model and show that planetary embryos can migrate inwards in a convoy. This is caused by the mutual interactions between the bodies, which increase their eccentricity. An increase in the planetary eccentricity quenches the entropy driven corotation torque \citep{2010A&A.523...A30} resulting in inward migration of these planets. As a consequence, multiple planetary embryos that grow exterior to the water ice line can migrate all the way to the inner edge in a convoy \citep{2019arXiv190208772I} in contrast to the single planetary embryos used in the here presented work. We thus think that the final semi-major axes of the planets in our simulations might be different if multi-body simulations are used, however, the growth of these embryos by pebble accretion is very fast and thus local, so that the water ice content is not affected much by the multiplicity \citep{2019arXiv190208772I}.

The planets that grow entirely exterior to the water ice line and are trapped until they reach the pebble isolation mass have the largest water ice content, as shown by \citet{2019A&A...624A.109B}. The water ice content of the planets that formed completely interior to the water ice line is by construction zero, while the water ice content is largest for planets formed just exterior to the water ice line. Planets forming even further away from the water ice line have a lower water ice content, because these planets might accrete CO$_2$, CH$_4$ and CO reducing the relative mass fraction of water ice. However, these planets would have a higher ice fraction. In fact the water ice content of planets formed just exterior to the water ice line and planets formed completely exterior to the CO ice line differ by about a factor of two. The planets formed far away from the star are dominated in CO$_2$, CH$_4$ and CO ice instead of water ice. As expected the water ice content of the planets is between the maximal allowed value (Fig.~\ref{fig:solarT100}) and zero.

Another factor that can influence the water ice content of the formed planets is the direction of migration. \citet{2019A&A...624A.109B} showed that planets have a lower water ice content in a scenario where planets can only migrate inwards. We see the same effect in our simulations here (see appendix~\ref{ap:formation}).

\subsection{Sub- and super-solar metallicity}

\begin{figure*}
 \centering
 \includegraphics[scale=0.7]{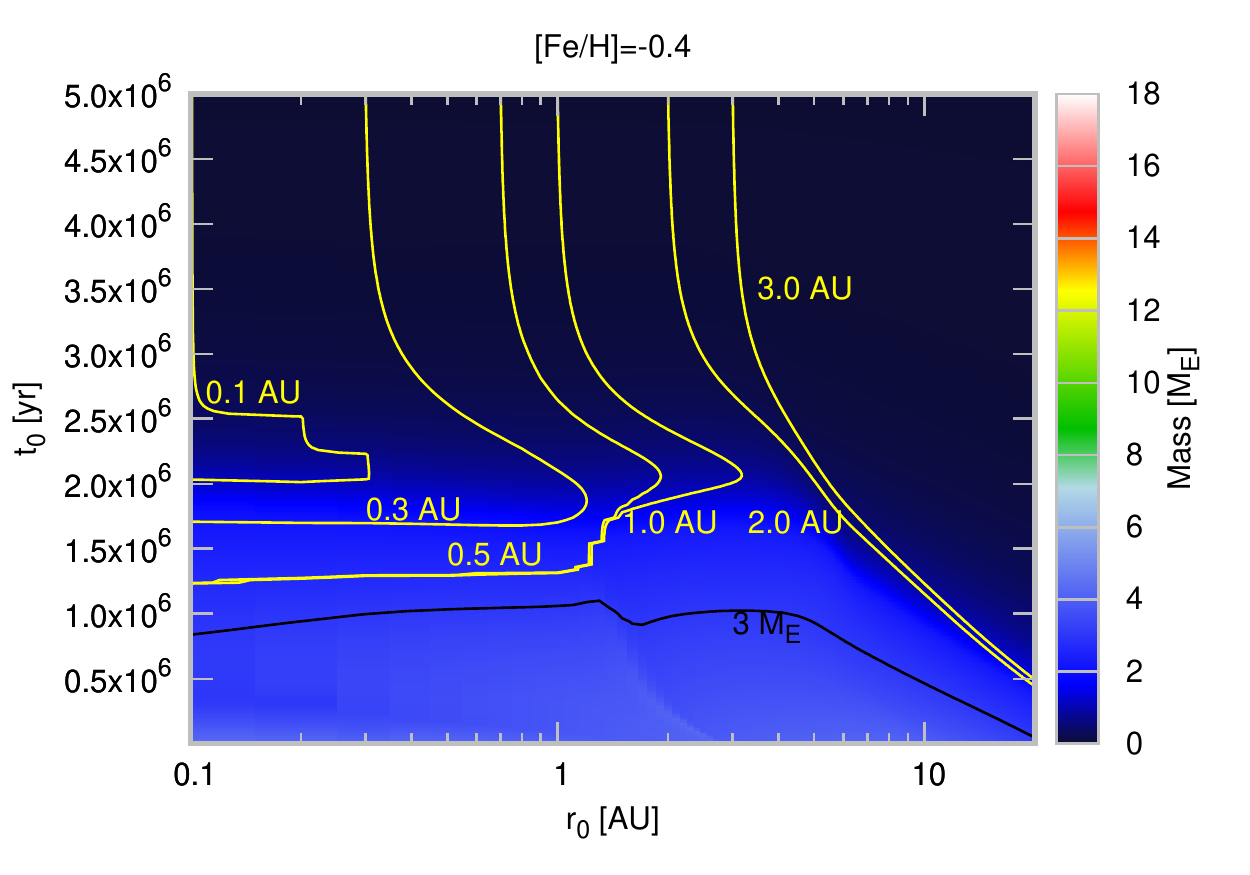} \quad
 \includegraphics[scale=0.7]{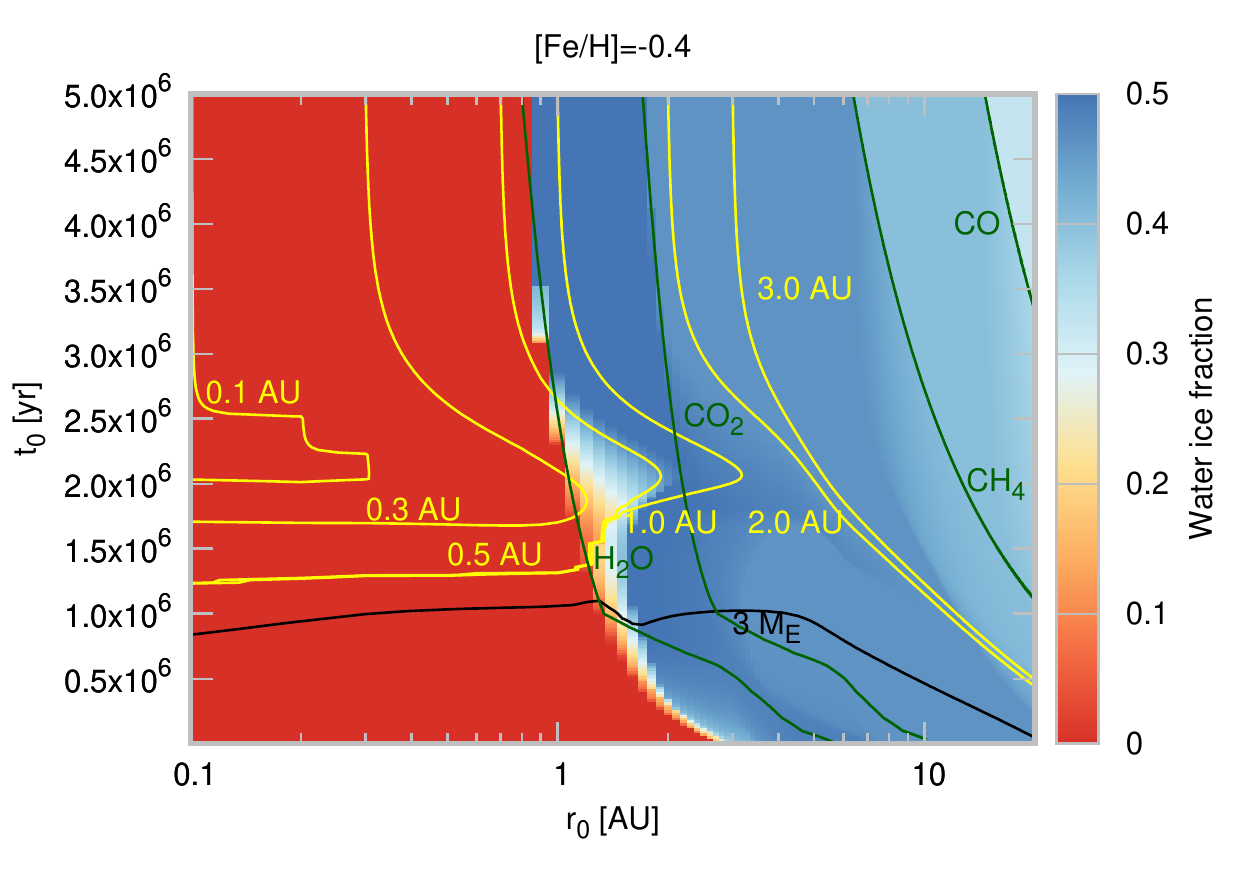}
 \includegraphics[scale=0.7]{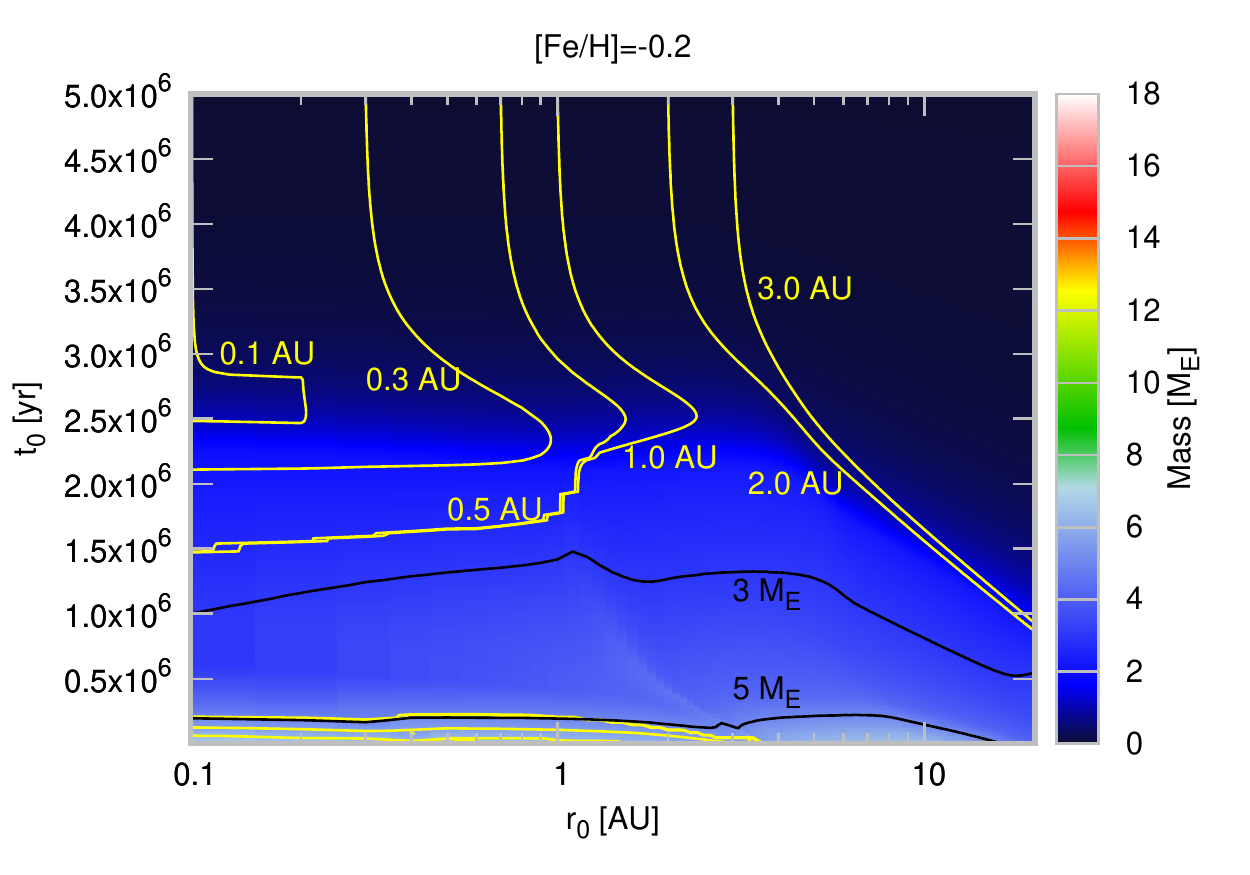} \quad
 \includegraphics[scale=0.7]{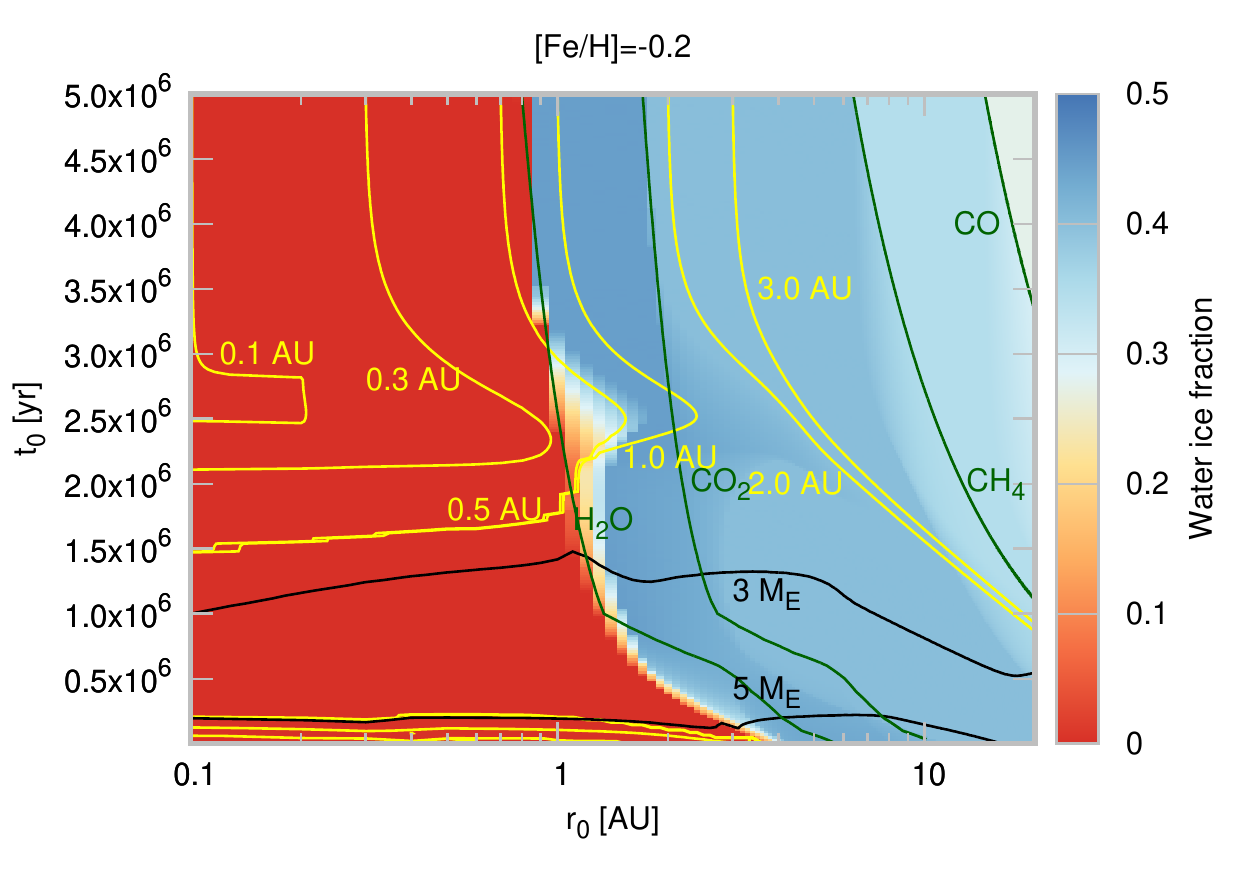}
 \includegraphics[scale=0.7]{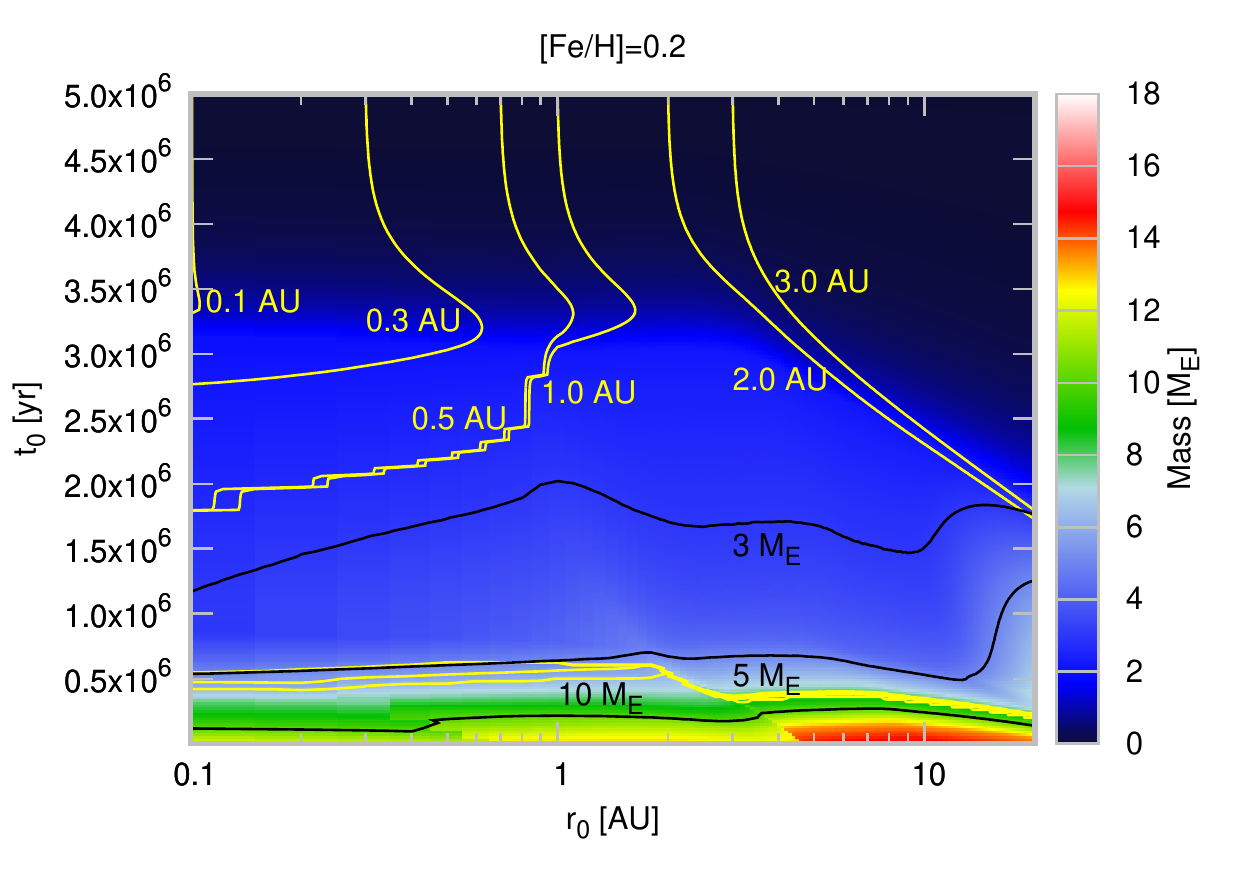} \quad
 \includegraphics[scale=0.7]{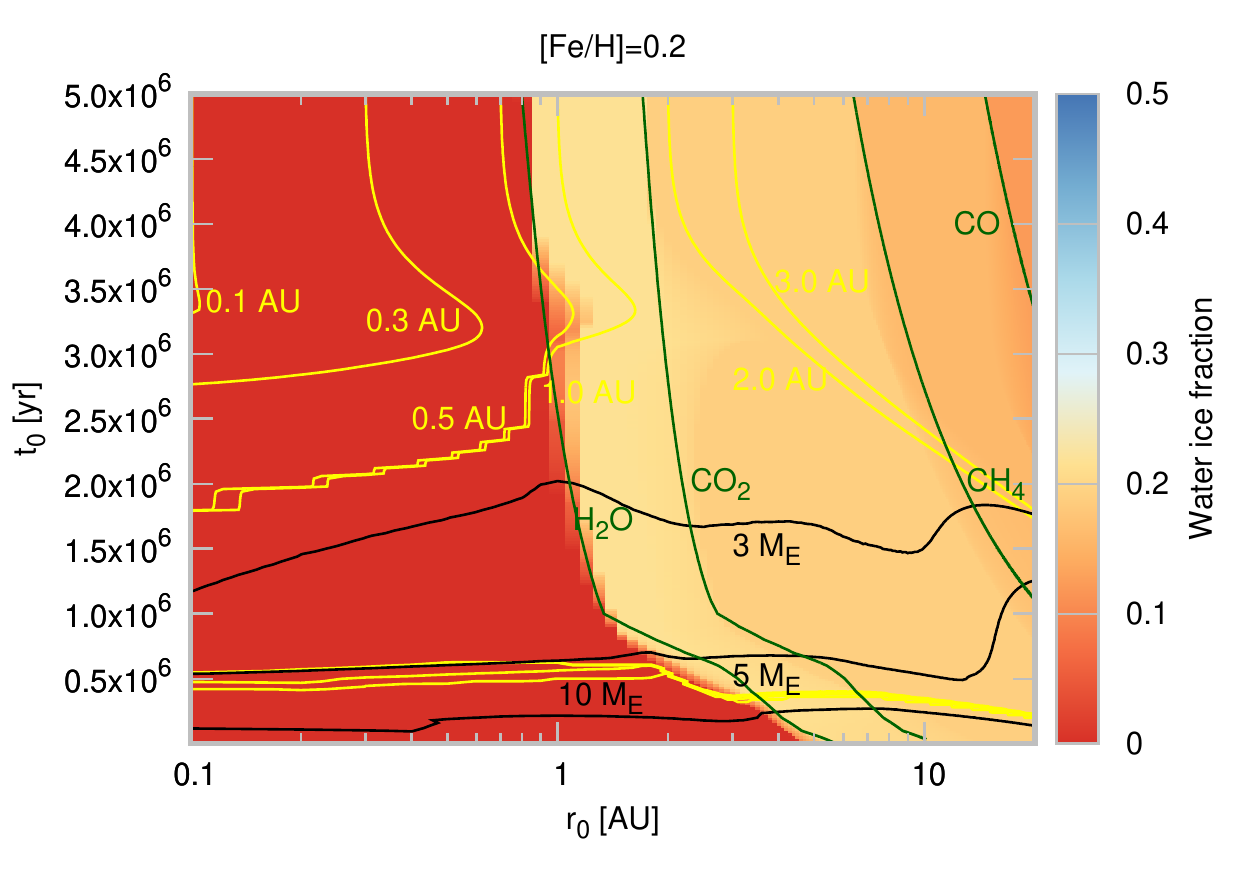}
 \includegraphics[scale=0.7]{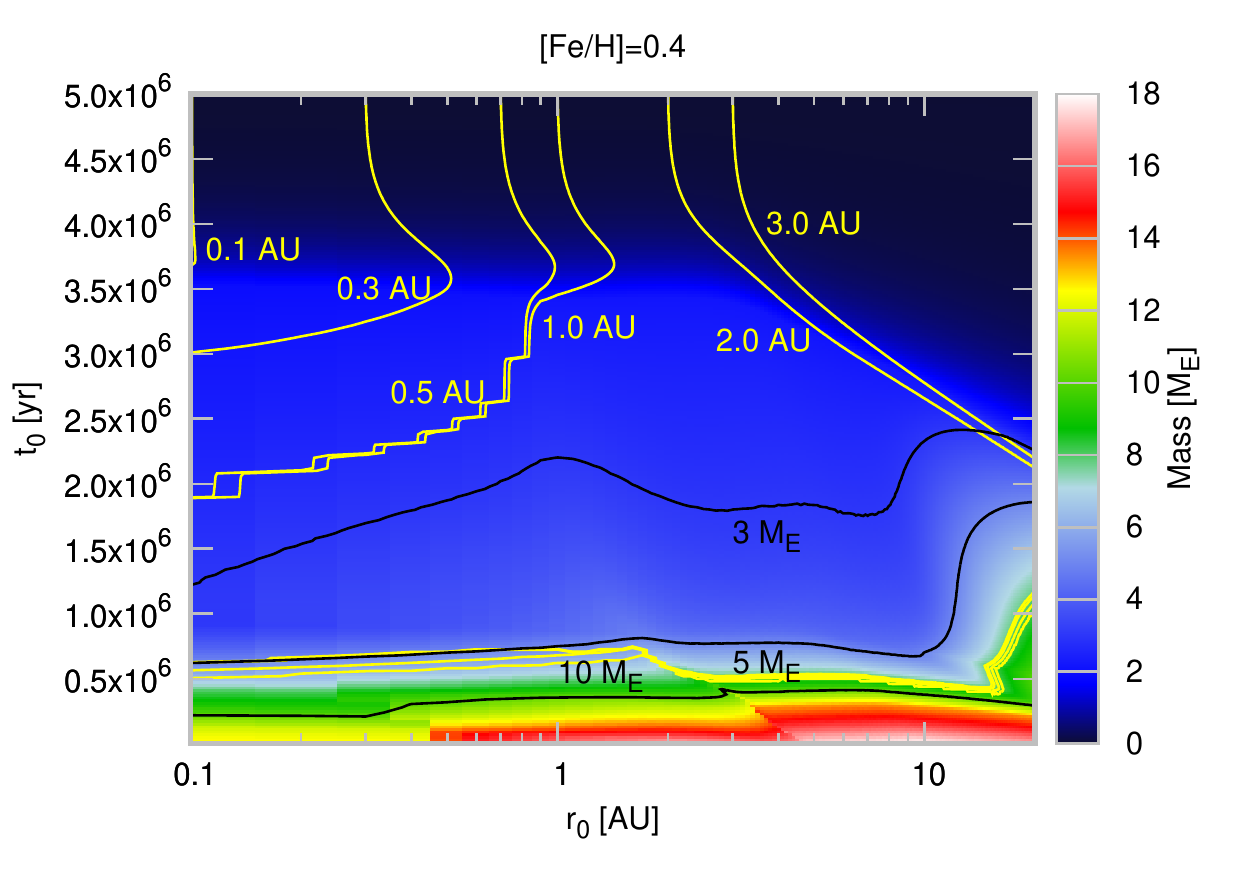} \quad
 \includegraphics[scale=0.7]{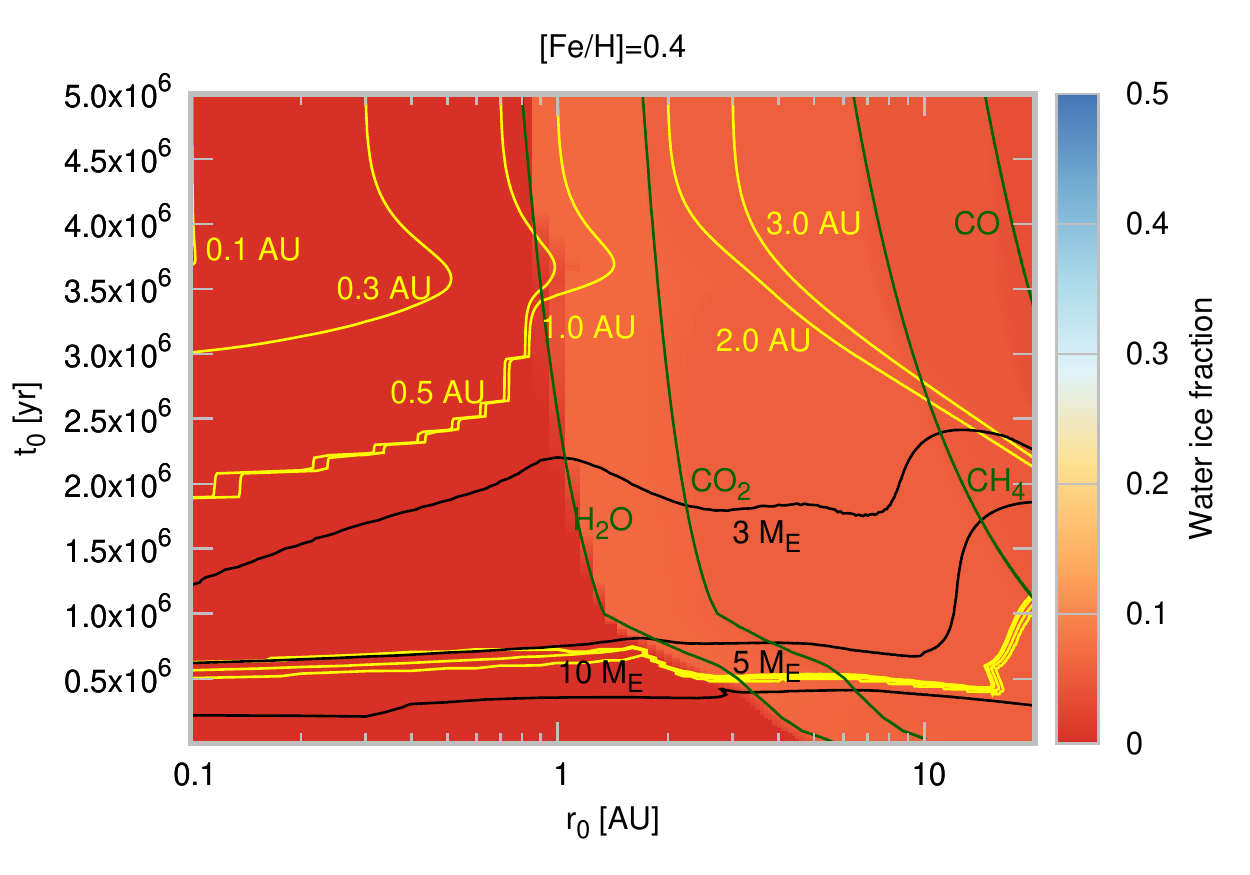} 
 \caption{Same meaning as Fig.~\ref{fig:PlanetsFeH0}, except that the planets are formed around stars with different [Fe/H] values and thus chemical abundances for the all elements (Fig.~\ref{fig:abundances}). Stellar abundances increase from top to bottom. As the stellar abundances increase, the water ice content of the planet decreases due to the lack of water ice, caused by the increasing C/O ratio (Fig.~\ref{fig:CO}).
    \label{fig:PlanetsFeHall}
   }
\end{figure*}

In Fig.~\ref{fig:PlanetsFeHall} we show the final masses, positions and water ice fractions of planets formed at different initial distances $r_0$ and at different initial times $t_0$ around stars with sub- and super-solar metallicity. The exact elemental abundances for each [Fe/H] value are shown in Fig.~\ref{fig:abundances} and table~\ref{tab:abu}.

Two effects for increasing host star metallicity are immediately clear from Fig.~\ref{fig:PlanetsFeHall}, (i) the planetary masses increase with increasing [Fe/H] and (ii) the water ice content of the planets decreases with increasing [Fe/H]. Both effects are not surprising. The increasing planetary masses with [Fe/H] are simply related to the higher pebble flux at larger [Fe/H], which allows faster accretion and as a result the planets can reach the pebble isolation mass earlier, when the pebble isolation mass is still large. The reduced water ice content with increasing [Fe/H] is simply related to the overall reduced water ice abundance at higher [Fe/H] from our chemical model.

There are, however, some other small differences for the planet formation pathway. In the low metallicity cases, the planets grow very slow, meaning that they just reach 2-3 Earth masses. Planets of that mass range are trapped in the region of outward migration until disc dissipation at 5 Myr and thus do not migrate to the inner edge. As a consequence at low [Fe/H], basically no planet reaches the inner edge in our disc model. However, chains of multiple planets can migrate inwards due to their mutual increase in eccentricity which reduces outward migration due to the entropy driven corotation torque \citep{2010A&A.523...A30, 2017MNRAS.470.1750I, 2019arXiv190208772I}. If the metallicity is slightly larger ([Fe/H]>-0.2), some planets become larger than 5 Earth masses, which allows them to migrate to the inner edge of the disc towards the end of the discs lifetime (Planets below the yellow lines at low $t_0$ in Fig.~\ref{fig:PlanetsFeHall} migrate all inwards to 0.1 AU).

As already shown in Fig.~\ref{fig:PlanetsFeH0}, the water ice content of the planets reduces if they are formed further away and the water ice content of the planets formed close to the water ice line is largest. This is, as stated before, caused by the fact that the planets formed exterior to the CO$_2$, CH$_4$ or CO ice line accrete these ices as well. As a result the water ice content of the planet decreases, but the total ice fraction can increase (not shown).

The total water ice content of the planets formed in the simulations is between the maximal amount of water allowed by the chemical model for a given stellar composition (see Fig.~\ref{fig:Moleculetrend}) and zero. Planetary migration across the water ice line can reduce the water ice content of growing planets, see appendix~\ref{ap:formation}. In particular if planetary migration is only directed inwards, the water ice content can be reduced compared to a scenario where planet migration is outwards at the water ice line \citep{2019A&A...624A.109B}. Nevertheless, the highest water ice content of planets can be achieved at low host star metallicity, due to the lowest C/O ratio, which allows the largest water ice content, independently if migration is inwards or outwards.

\section{Discussion}
\label{sec:disc}

\begin{figure}
 \centering
 \includegraphics[scale=0.7]{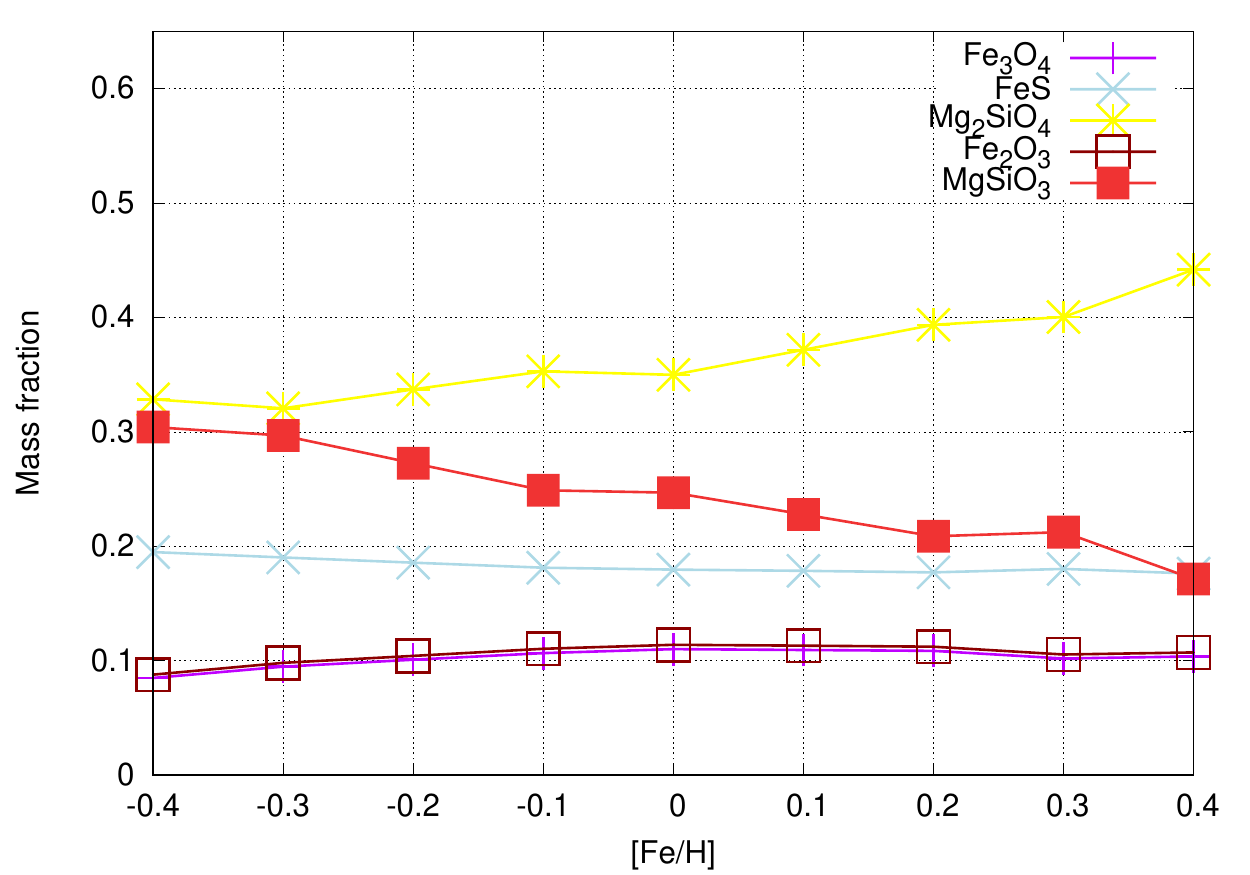}
 \includegraphics[scale=0.7]{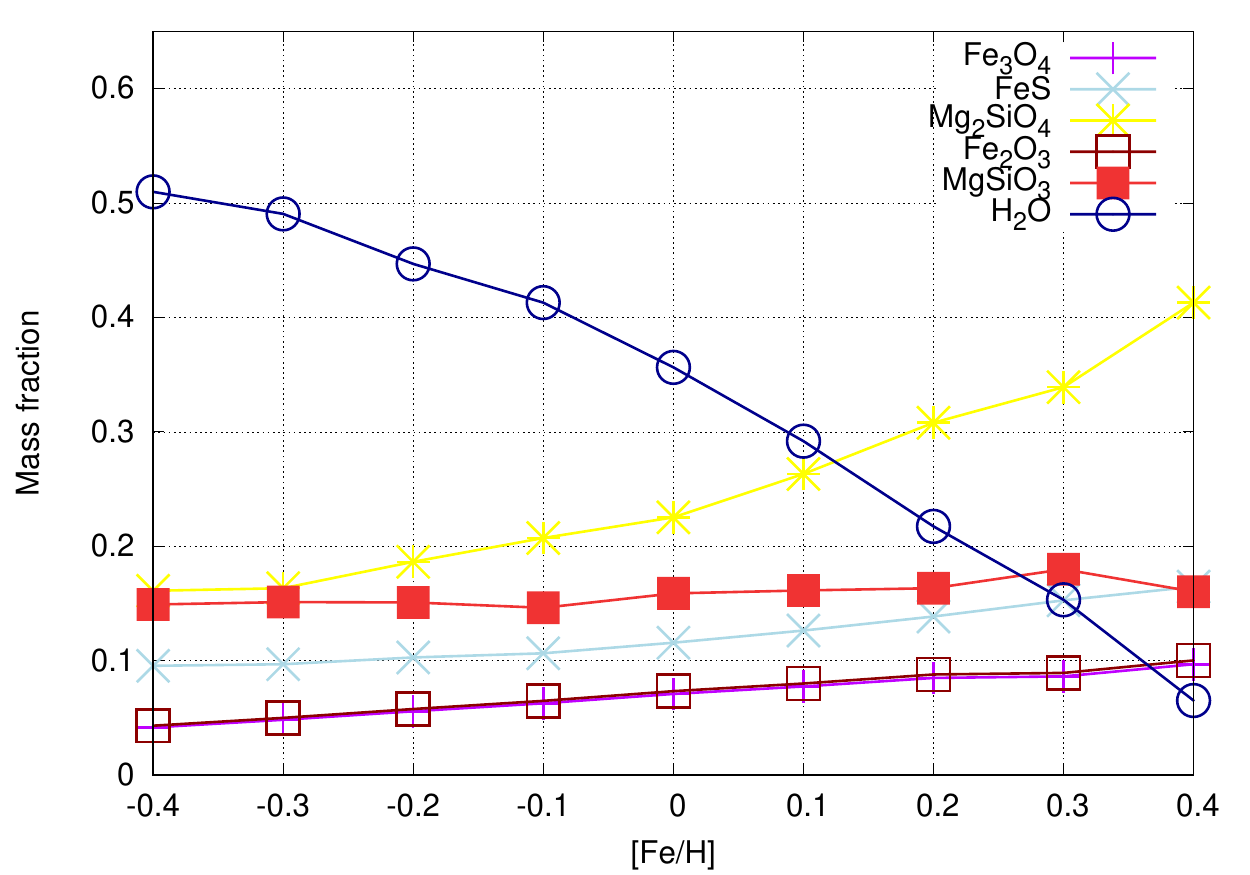}
 \caption{Molecular mass fractions of solid planetary building blocks formed completely interior (top) to the water ice line and completely exterior (bottom) to the water ice line as function of [Fe/H]. The lines of Fe$_2$O$_3$ and MgSiO$_3$ are overlapping, because the mass fraction of these molecules is very similar. For solids formed interior to the water ice line, a clear rise in the Mg$_2$SiO$_4$ fraction and decline in the MgSiO$_3$ fraction is visible, caused by the increase of Mg/Si for increasing [Fe/H]. Solid planetary building blocks formed exterior of the water ice line show a clear trend of reduced water ice fraction with increasing [Fe/H], which is then compensated by an increase in Mg$_2$SiO$_4$.
    \label{fig:Moleculetrend}
   }
\end{figure}

\begin{figure}
 \centering
 \includegraphics[scale=0.7]{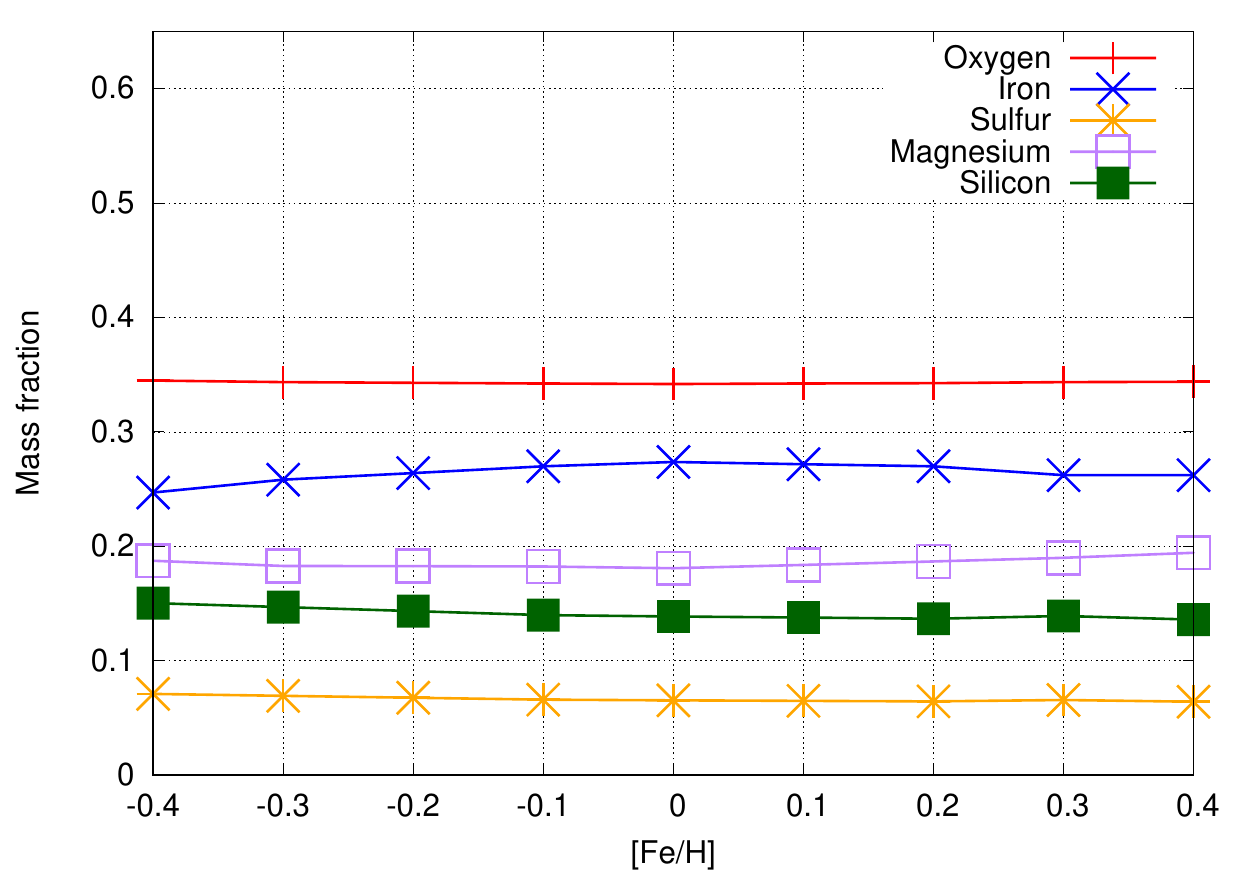}
 \includegraphics[scale=0.7]{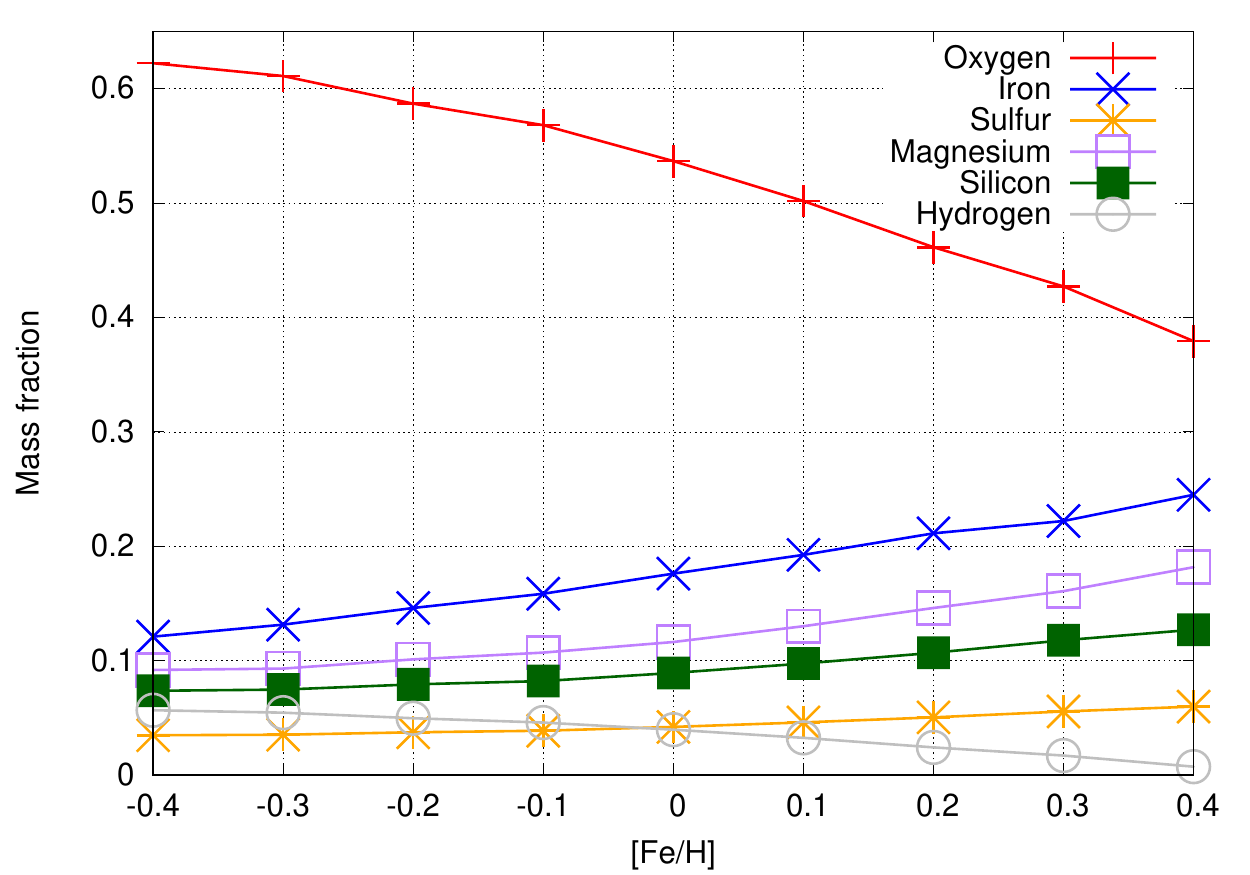}
 \caption{Elemental mass fractions of solid planetary building blocks formed completely interior (top) to the water ice line and completely exterior (bottom) to the water ice line as function of [Fe/H]. The elemental abundances of solids formed interior to the water ice line (T>150K) are roughly constant for all values of [Fe/H]. In contrast, the oxygen content inside the solids severely decreases for solids formed exterior to the water ice line (T<150K) due to the reduced oxygen abundances relative to C, Si, Mg, Fe and S for increasing [Fe/H].
    \label{fig:Elementtrend}
   }
\end{figure}

\begin{figure}
 \centering
 \includegraphics[scale=0.45]{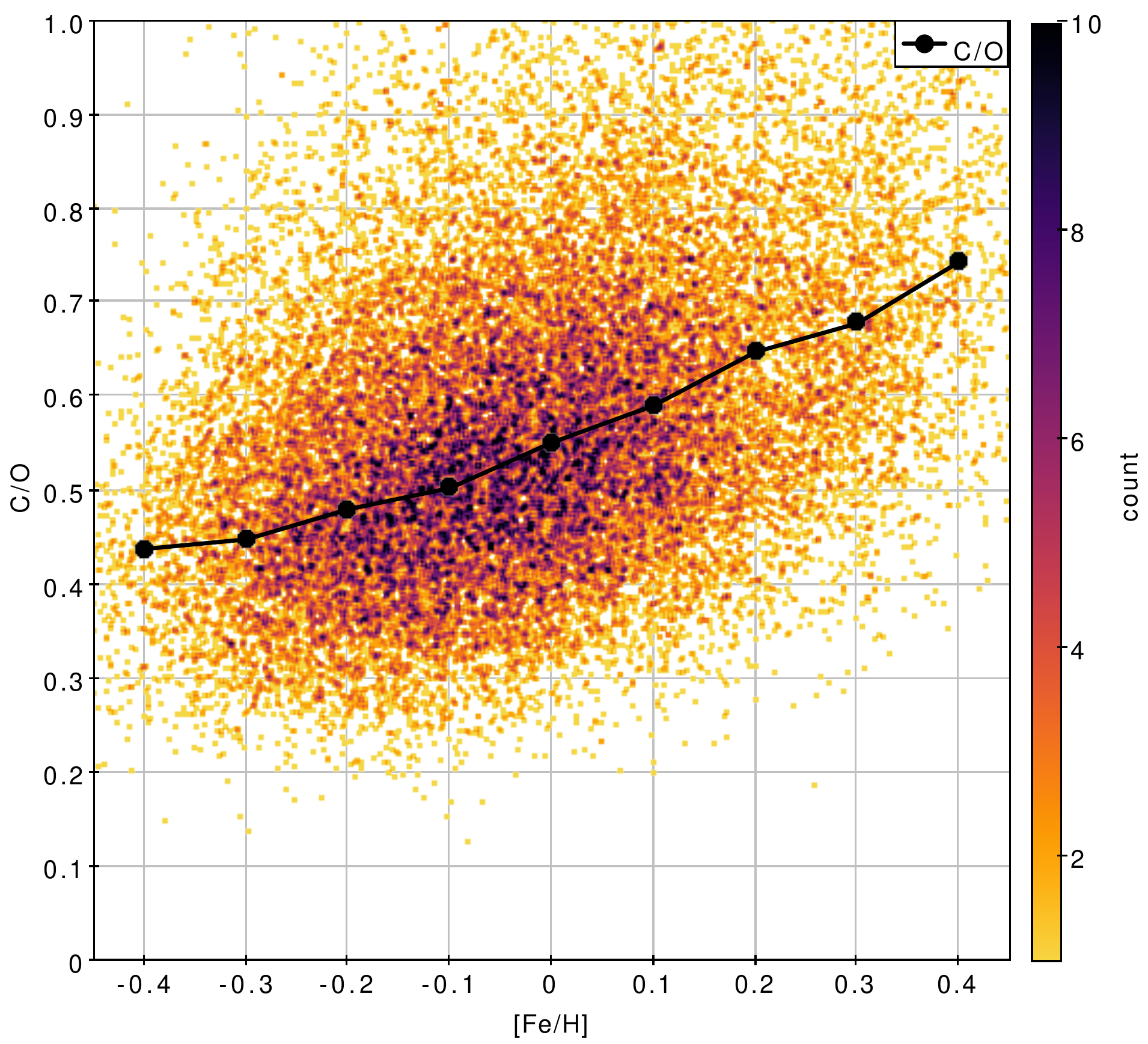}
 \caption{C/O ratio as function of [Fe/H] with the stars containing good carbon and oxygen values from the GALAH survey. The black line is originating from the data shown in Fig.~\ref{fig:abundances} and marks the mean C/O ratio from all stars in our sample. The number of the stars shown here is less than used to derive the [C/H] and [O/H] value in Fig.~\ref{fig:abundances}, because stars with good carbon value might not have a good oxygen value at the same time. As [Fe/H] increases, so does the C/O ratio. This implies that more oxygen can be bound with carbon to form CO and CO$_2$ for increasing [Fe/H].
    \label{fig:CO}
   }
\end{figure}

In Fig.~\ref{fig:Moleculetrend} and Fig.~\ref{fig:Elementtrend} we show the molecular and elemental mass fractions of solid planetary building blocks formed interior (T>150K) and exterior (T<150K) to the water ice line as function of [Fe/H]. These figures summarize the trends described in section~\ref{sec:placomp} to give a better overview over the trends. In addition we show in Fig.~\ref{fig:CO} the C/O ratio as function of [Fe/H]. In the following we discuss the assumptions and implications of the results of our model on planet formation, where we also show the results of a planet formation model in appendix~\ref{ap:formation}.

\subsection{Planet formation}

One of the main questions in planet formation theories is where the first planetesimals and planetary embryos formed that then grew to fully grown planets. In recent years, the water ice line has become the prime suspect for the first planetesimals to form \citep{2016A&A...596L...3I, 2017A&A...608A..92D}, also because of the large water abundance in the solar system \citep{2003ApJ...591.1220L}. Condensation of water onto grains allows them grow larger than with pure coagulation \citep{2013A&A...552A.137R, 2017A&A...602A..21S, 2019arXiv190708471R}, which allows planetesimals to form easier. This is caused by the larger particle sizes which allow gravitational collapses of pebble clouds to planetesimals at lower overall metallicities \citep{2017A&A...606A..80Y}. In contrast, the CO ice line is not thought to harbour the same effect as the water ice line, because the grains drift inwards faster than they can grow through condensation at these distances \citep{2017A&A...600A.140S}.

However all these simulations have been conducted with solar abundances of water and CO. As the host star metallicity increases, our model predicts a significant reduction in water (Fig.~\ref{fig:Moleculetrend} and an increase in CO (Fig.~\ref{fig:CO})). A change in the water ice fraction has large implications for the formation of planets around the water ice line \citep{2016A&A...590A.101B}.

In the case of low host star metallicity, the water ice fraction is much larger than for solar metallicity, which could increase the effects of condensation and could thus allow grains to grow larger. This could significantly help planet formation around the ice line in low metallicity environments, where building blocks are rare, but the increased pebble sizes due to condensation allow an easier collapse of pebble clouds to planetesimals \citep{2017A&A...606A..80Y}. However, this effect might be counteracted by the overall low metallicity which hinders the gravitational collapse of pebble clouds to planetesimals as well \citep{2017A&A...606A..80Y}. This implies that the effect of condensation can not be so strong that giant planet cores grow quickly and easily, due to the lack of giant planets around low metallicity host stars \citep{2004A&A...415.1153S, 2005ApJ...622.1102F, J2010}. Additionally, it could imply that super-Earths formed around low metallicity stars could be water rich, in contrast to super-Earths formed around high metallicity stars, where water ice is rare.

On the other hand, the water ice fraction is very low around stars with super-solar metallicity, which implies that maybe condensation is not the main driver of grain growth and thus planetesimal formation around these stars. Condensation at the water ice line is reduced, but it could be compensated by the overall increased metallicity around these stars, which can trigger planetesimal formation (see \citealt{Johansen2014} for a review). The simulations of \citet{2017A&A...606A..80Y} show that even small particles with Stokesnumbers around $10^{-3}$ can trigger planetesimal formation, if the overall metallicity is super-solar, indicating that the help of condensation to particle growth is not necessarily required to form planetesimals.

The giant planet occurrence rate increases with host star metallicity \citep{2004A&A...415.1153S, 2005ApJ...622.1102F, J2010}, which is confirmed by many simulations in the core accretion scenario \citep{2004ApJ...604..388I, 2009A&A...501.1139M, 2017ASSL..445..339B, 2018MNRAS.474..886N}. However, these simulations did not take the effects of different elemental scalings with host star metallicity into account and should be revisited due to the effects found here. In addition, the amount of CO and CO$_2$ around these stars is increased compared to stars with solar abundances due to their elevated C/O ratio (Fig.~\ref{fig:CO}). This could allow condensation at the CO snow line to outperform radial drift, implying that giant planet formation around stars with solar and sub-solar metallicity starts at the water ice line where condensation can increase the pebble sizes, while giant planet formation around high metallicity stars could start at the CO ice line, where our model predicts a larger amount of CO available for for condensation compared to solar and sub-solar metallicity. This, however, needs to be tested with future models like of \citet{2017A&A...608A..92D} that combine grain growth and drift with condensation and planetesimal formation. In addition the C/O ratio of giant planet host stars could help to distinguish if water ice condensation is the main driver of giant planet formation.

\subsection{Composition of super-Earths and gas giants for different host star abundances}

In our model we study the composition of solid material around host stars with different metallicities. We use this to calculate the composition of solid planetary building blocks that form interior or exterior to the water ice line. In section~\ref{sec:formation} we show the water ice content of planets using a planet formation model featuring planet migration, disc evolution and pebble accretion, which resulted in water ice contents of the formed planets between zero and the maximal content allowed for the specific host star abundance. In appendix~\ref{ap:formation} we explain our planet formation model in detail. Our planet formation simulations coupled to our chemical model shows that the water ice content of the super-Earths is between zero and the maximal allowed water content of the model (Fig.~\ref{fig:PlanetsFeHall} and Fig.~\ref{fig:Moleculetrend}). This result is consistent with other simulations of planet formation including compositions and planet migration \citep{2017MNRAS.464..428A, 2019A&A...624A.109B}.

In Fig.~\ref{fig:Moleculetrend} we show the mass fraction of the molecules that contribute to solid planetary building blocks for changing host star metallicities. For planets forming completely interior to the water ice line (T>150K), the molecular composition of the solids only changes slightly, with an increase of Mg$_2$SiO$_4$ by about 15\% from [Fe/H]=-0.4 to 0.4, while MgSiO$_3$ decreases by the same amount over the same range of [Fe/H]. This is caused by the relative stronger increase of magnesium compared to silicon with [Fe/H], see Fig.~\ref{fig:abundances}. In the elemental mass fraction of the solids (Fig.~\ref{fig:Elementtrend}), on the other hand, the relative mass fractions of all elements stay roughly constant with [Fe/H]. This is caused by a very similar oxygen and magnesium fraction inside of MgSiO$_3$ and Mg$_2$SiO$_4$.

This implies that the overall composition of rocky planets that formed completely interior to the water ice line (T>150K) is very similar for stars with different metallicity. For rocky planets at solar composition our simulations actually predict a composition that is very close to the Earth \citep{McDonough95}. On the other hand, some super-Earths have densities much higher than the Earth \citep{2017A&A...608A..93G}, similar to Mercury. These high densities are caused by a large iron fraction, which we do not reach in our model. For Mercury, the high density could be explained by stripping its mantle through a collision, leaving an iron rich remnant \citep{1988Icar...74..516B}. In simulations that study the formation of close-in super-Earths, collisions are basically inevitable \citep{2015A&A...578A..36O, 2017MNRAS.470.1750I, 2018A&A...615A..63O, 2019arXiv190208772I, 2019arXiv190208694L}, but collisions are normally treated as perfect mergers in these simulations and they thus do not account for mantle stripping. However, this process could help to explain the large densities of some super-Earth planets.

Planets formed completely exterior to the water ice line (70K<T<150K), on the other hand, would show a strong decrease in water ice with increasing host star metallicity (Fig.~\ref{fig:Moleculetrend} and Fig.~\ref{fig:H2Ocontent}). This is caused by the binding of oxygen with carbon to form CO and CO$_2$ and the other rock forming elements, which results in only a small fraction of oxygen that can form water ice. As water ice decreases, Mg$_2$SiO$_4$ increases in a similar fashion as for planets formed interior to the water ice line. The elemental trend here gives a clear picture as well (Fig.~\ref{fig:Elementtrend}). As the water ice decreases, the oxygen mass fraction decreases and the mass fraction of the other elements increases accordingly.

The consequences for the composition are quite dramatic and super-Earths formed exterior to the water ice line have a very significant water ice fraction if they form around low metallicity stars and a very tiny water ice fraction if they form around high metallicity stars (Fig.~\ref{fig:PlanetsFeHall}). This has important consequences for the formation mechanism of close in super-Earths.

In the Kepler sample, recent detailed observations have revealed a gap in the radii of the observed super-Earths \citep{2017AJ....154..109F, 2018MNRAS.479.4786V}. This gap is generally interpreted as atmospheric mass loss of small planets either by photoevaporation \citep{2013ApJ...775..105O, 2014ApJ...792....1L} or by mass loss directly from the cooling of the core \citep{2019MNRAS.487...24G}. As a consequence, it is thought that most close-in super-Earths are of mostly rocky nature \citep{2017ApJ...847...29O, 2018ApJ...853..163J}, putting constraints on planet formation models. However, some models predict that these super-Earths could have water contents up to 10-20\% \citep{2019MNRAS.487...24G}, in agreement with recent observations \citep{2019arXiv190604253Z}.

For example, the N-body simulations of \citet{2019arXiv190208772I} show that planets assembled exterior to the water ice line should have a large water ice content (50\% in their model), mostly inconsistent with the observations. On the other hand, if the assembly of the planets starts exterior to the water ice line, but finishes interior to the water ice line, the final composition could be in line with the observations \citep{2018MNRAS.479L..81R, 2019A&A...624A.109B}. For super-Earths formed around high metallicity stars with [Fe/H]>0.2, our model predicts a maximum water ice content of 20\%, which is naturally consistent with the observed gap in the planetary radius distribution. This could imply that the radius gap could only tell something about the formation channel of super-Earths around stars with [Fe/H]<0.2. Thus a more detailed analysis of the host stars within the Kepler sample is of crucial importance to constrain planet formation theories.

We apply our model to the solid building blocks of planets and have not focused on the gaseous component of the protoplanetary discs and what this could imply for the composition of giant planets. Giant planets that grow and migrate in protoplanetary discs do not only accrete hydrogen and helium in the gas phase, but also heavy elements. The accretion of these elements can be enhanced due to the evaporation of volatiles (e.g. H$_2$O, CO$_2$, CO) at ice lines \citep{2011ApJ...743L..16O, 2017MNRAS.469.4102M, 2017MNRAS.469.3994B}. As already mentioned before, the C/O ratio in the gas phase changes with host star metallicity (Fig.~\ref{fig:abundances} and Fig.~\ref{fig:CO}). In addition, hot Jupiters, which are the gas giants whose atmosphere we can characterize, are more common around high metallicity stars \citep{2018arXiv180206794B}. This could naturally imply a change in the C/O ratio of the observed giant planet atmospheres compared to solar composition and should be taken into account in the modeling of the planet formation pathway of these planets, where previous simulations were not tailored to match the exact host star abundances \citep{2017MNRAS.469.4102M, 2017MNRAS.467.2845A}.

\subsection{Habitability}

Planets that are suspect to harbour life, as we know it, reside in the habitable zone around their host star (e.g. \citealt{2013Sci...340..577S}), which is defined as a region around the star where water could be in liquid form\footnote{Life based on other molecules than H$_2$O could have different habitable zones \citep{2017arXiv170310803T}.}. This implies that a planet inside the habitable zone should actually harbour water. In the solar system, the Earth formed dry in the inner regions of the protoplanetary disc and water was delivered via impacts of asteroids and comets triggered by the giant planets \citep{2009Icar..203..644R, 2017Icar..297..134R}. Simulations normally assume a water ice composition gradient around the water ice line, where the most water rich objects can have a substantial amount of water ice. However, around stars with high metallicity, the water ice fraction is dramatically reduced (Fig.~\ref{fig:Moleculetrend}), which could have important implications for the potential habitability of planets inside the habitable zone around high metallicity stars. The study by \citet{2012ApJ...756..178A}, on the other hand, indicates that too much water could actually be potentially dangerous for the habitability of a given planet, implying that planets around stars with a lower metallicity might be more suitable for life. However, as already hinted above, water delivery onto planets is a complicated process that needs much further investigation. Nevertheless, our model suggests that the characterization of potential habitable worlds with future instruments should take the host star composition into account.

\subsection{Chemical Model}

Our used chemical model (table~\ref{tab:species}) is a simple equilibrium model that does not take the chemical evolution during the protoplanetary disc stage into account, which can alter the chemical abundances of different molecules \citep{2013ChRv..113.9016H, 2016arXiv160706710E}. Our used model also does not account for the formation of molecules during the infall stage, however simulations using solar-like composition \citep{2015A&A...584A.124F} seem to be in agreement with our model at solar composition. However, the impact of a more advanced model seems to be much more profound on the gaseous component \citep{2018A&A...613A..14E} and thus the planetary atmosphere composition \citep{2019A&A...627A.127C} than on the solid composition. However, to calculate the solid composition of the material, also the gaseous component has to be taken into account. This is especially important for the oxygen fraction, as a large fraction of oxygen is bound in CO and CO$_2$, especially for high C/O ratios.

In the study by \citet{2017A&A...608A..94S} the effect of gas phase CO and CO$_2$ was not taken into account and they find that the protoplanetary discs should harbour a large water fraction independently of the host star metallicity. We attribute this difference in their model, originally presented in \citet{2015A&A...580L..13S}, to ours to the more detailed chemical model we use here. In particular we include the effects of oxygen binding in CO and CO$_2$, which binds most of the oxygen, especially for high metallicity stars with large C/O ratio (Fig.~\ref{fig:CO}). Additionally, we include binding of oxygen with iron to form Fe$_3$O$_4$ and Fe$_2$O$_3$. This results in the low water ice fraction around high metallicity stars in our model compared to \citet{2017A&A...608A..94S}.

As the water ice fraction is closely related to the binding of oxygen with carbon to form CO and CO$_2$, the water ice fraction would change if a lot of carbon grains would be bound in molecules that do not feature any oxygen, for example in organics or CH$_4$ (which accounts for 45\% of the carbon abundance in our nominal model). We present a chemical model in appendix~\ref{ap:carbon}, where we decrease the CO abundance to 0.25 of the C/H and instead include an organic component consisting of pure carbon grains with evaporation temperature of 500K \citep{2003A&A...410..611S}, but leaving the other parts as in our nominal model (table~\ref{tab:species}). We find that the water ice content at [Fe/H]=0.4 can increase to 15\% for solid planetary building blocks formed exterior to the water ice line. However this is still a factor of 3-4 smaller than the water ice content at [Fe/H]=-0.4, reproducing the trend shown in Fig.~\ref{fig:Moleculetrend} from our nominal model. As the water ice content of planets formed exterior to the water ice line is crucial for planet formation, planetary composition and even habitability, future studies of planet formation around stars with different metallicity should take even more detailed chemical models into account. However, some models (e.g. \citealt{2018A&A...613A..14E}) seem to indicate that carbon is most abundant in the form of CO or CO$_2$ rather than organic material without any oxygen, indicating a trend more consistent with our nominal chemical model (Fig.~\ref{fig:Moleculetrend}).

\subsection{Future observations}

In the near future GALAH will publish its third data release. This release will contain more data collected after the DR2 but also re-analysis of the previous stars: this will include abundances of new elements (some neutron-capture elements not treated in DR2) and might incorporate new non-LTE corrections coming from 3D models, which could influence the here drawn conclusions.

In addition to this, in the next years other big spectroscopic surveys will start their operation. In particular 4MOST (4m Multi-Object Spectroscopic Telescope, more information on all the surveys can be found in \citet{2019Msngr.175....3D}) will start to observe at the end of 2021 from the VISTA telescope in Chile and it is planned to observe almost 10 million stars in the Milky Way. Related to this star-planet connection, more than 2.5 millions of objects will be observed in the Galactic disk, both in high (R $\sim$ 20,000) and low resolution (R $\sim$ 5000): in particular for the high resolution disk and bulge survey, at least 15\% of the targets will be observed also by TESS. For the high-resolution observations more than 20 elemental abundances can be potentially derived and in this list all the elements that are included in this work are going to be observed, including sulfur. The potential coming from 4MOST is incredible: the massive sample homogeneously analysed together with the big overlap with the planet transit mission TESS and other surveys will give enormous amount of constrains and data for planet formation theory, in particular in respect to the star-planet connection.

\section{Summary}
\label{sec:summary}

In this study we have combined observations of elemental abundances of stars (Fig.~\ref{fig:abundances}) to the composition of planet forming material and planet formation. Our model is based on the assumption that the host star abundance is directly linked to the abundances of elements within the planet forming protoplanetary disc. To calculate the exact composition of solid planetary building blocks interior and exterior to the water ice line we have used a simple equilibrium chemical model (table~\ref{tab:species}) and show variations of that model in appendix~\ref{ap:carbon}.

Our model predicts that the elemental composition of planetary building blocks formed completely interior to the water ice line (T>150K) is very similar across all host star metallicities (Fig.~\ref{fig:Elementtrend}). However, the elemental mass fractions of planetary building blocks forming completely exterior to the water ice line (T<150K) are very different depending on the host star metallicity. In particular the water ice mass fraction reduces from close to 50\% at [Fe/H]=-0.4 to only $\sim$6\% at [Fe/H]=0.4 (Fig.~\ref{fig:Moleculetrend}) within our chemical model. This has important consequences for the formation, composition and habitability of planets.

Planet formation models (section~\ref{sec:formation} and appendix~\ref{ap:formation}) show that the water ice content of the formed planets can be mixed between the extreme states of zero water ice and the maximal content allowed for the specific disc composition due to planetary migration and the evolution of the water ice line (see also \citealt{2019A&A...624A.109B}). This trend has also been found in other planet formation simulations just using solar composition (e.g. \citealt{2017MNRAS.464..428A}).

Additionally, a change in the water ice fraction alters the formation pathway of planets \citep{2016A&A...590A.101B}. In particular, \citet{2016A&A...590A.101B} showed that discs with similar metallicity, but higher water ice fraction form more giant planets. Our here presented simulations show that the water ice fraction around high metallicity stars is very low and thus condensation of water ice molecules to assist grains growth \citep{2019arXiv190708471R} might not be the main reason to trigger giant planet formation. Instead planetesimal formation might be triggered by the overall large metallicity \citep{2017A&A...606A..80Y}. The effects of water ice condensation for grain growth might thus be more important for stars with [Fe/H]<0.2, while its effect might be very small around stars with even higher metallicity. We thus speculate that giant planet formation around stars with [Fe/H]<0.2 might be triggered at the water ice line, but it could be triggered instead at the CO ice line for stars with even higher metallicity due to the large CO abundance caused by the large C/O ratio (Fig.~\ref{fig:CO}), which could cause grain growth at the CO line due to CO condensation.

The change of the water ice fraction with host star metallicity has also direct impacts on the habitability of planets formed in these systems. Systems with intrinsically lower or no water abundance might not be able to form planets with enough water for them to be habitable. Future characterization of atmospheres of exoplanets in the habitable zone should thus take the host star abundances into account.

Our study clearly indicates a large impact on the planetary composition and planet formation pathway depending on the chemical composition of the host star. We thus encourage future studies of planet formation that try to span a wide range of host star metallicities to include the effects of different abundances of different elements.

\begin{acknowledgements}

B.B. thanks the European Research Council (ERC Starting Grant 757448-PAMDORA) for their financial support. C.B.'s work was supported by Sonderforschungsbereich SFB 881 "The Milky Way System" (subproject A9) of the German Research Foundation (DFG). We thank the referee for her/his comments.

\end{acknowledgements}

\appendix
\section{Stellar abundances}
\label{sec:abu}

\begin{figure}
 \centering
 \includegraphics[scale=0.5]{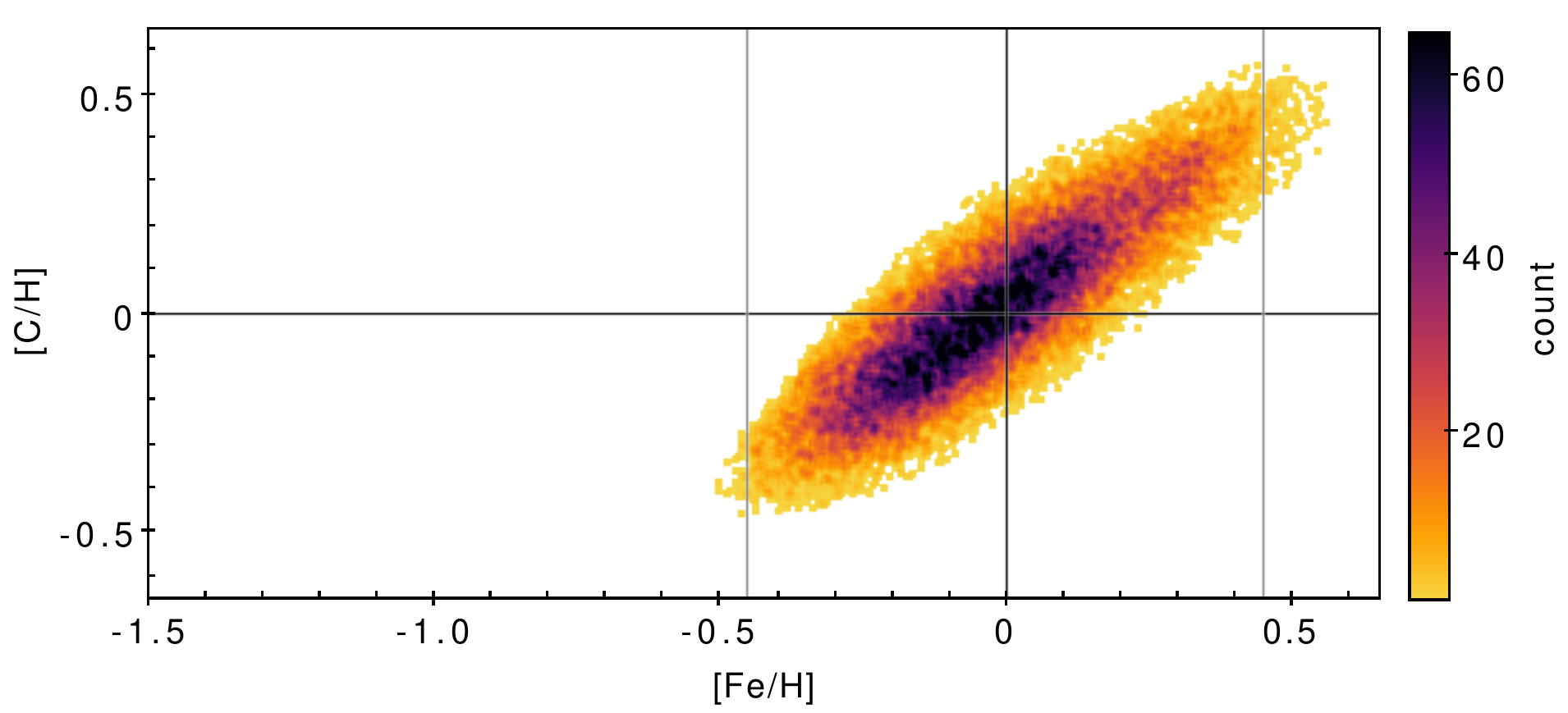}
 \includegraphics[scale=0.5]{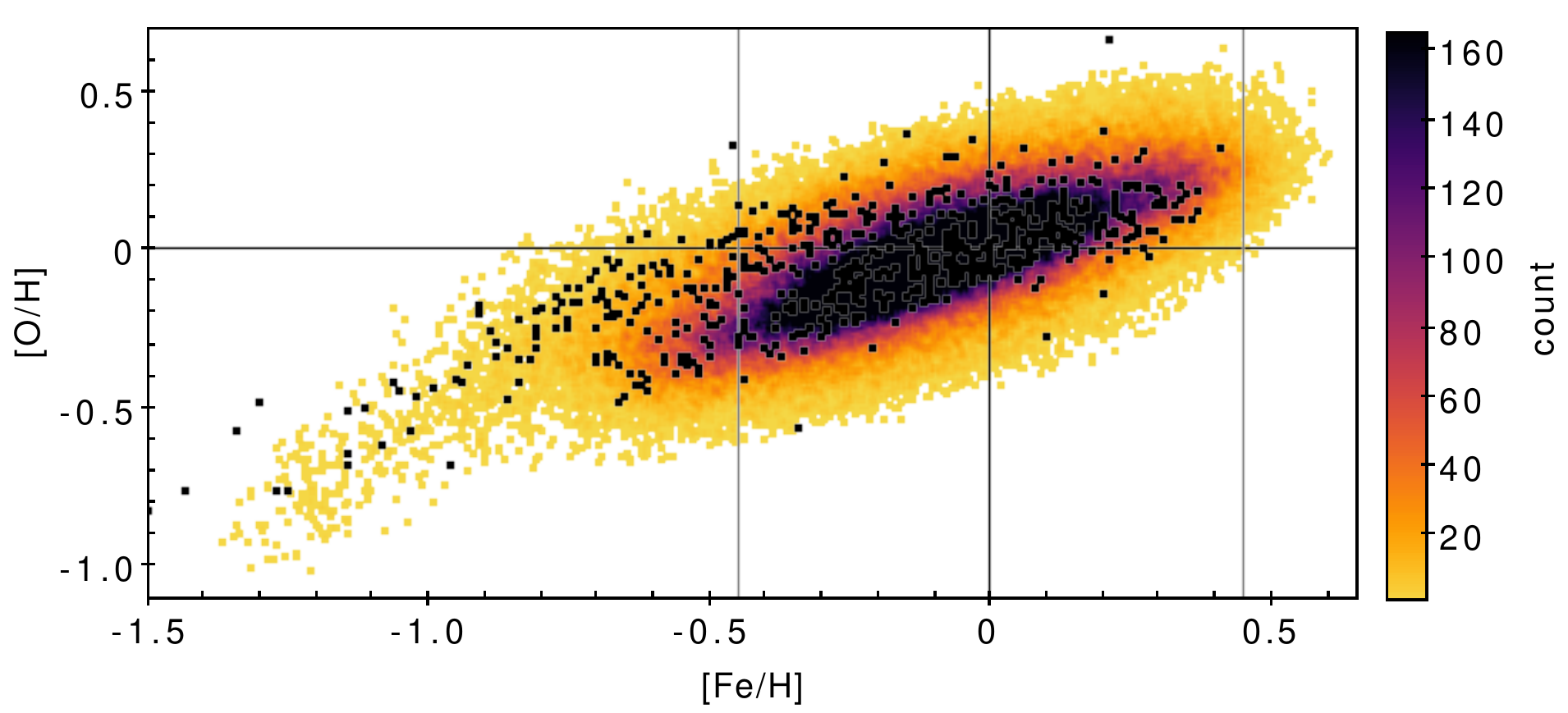}
 \includegraphics[scale=0.5]{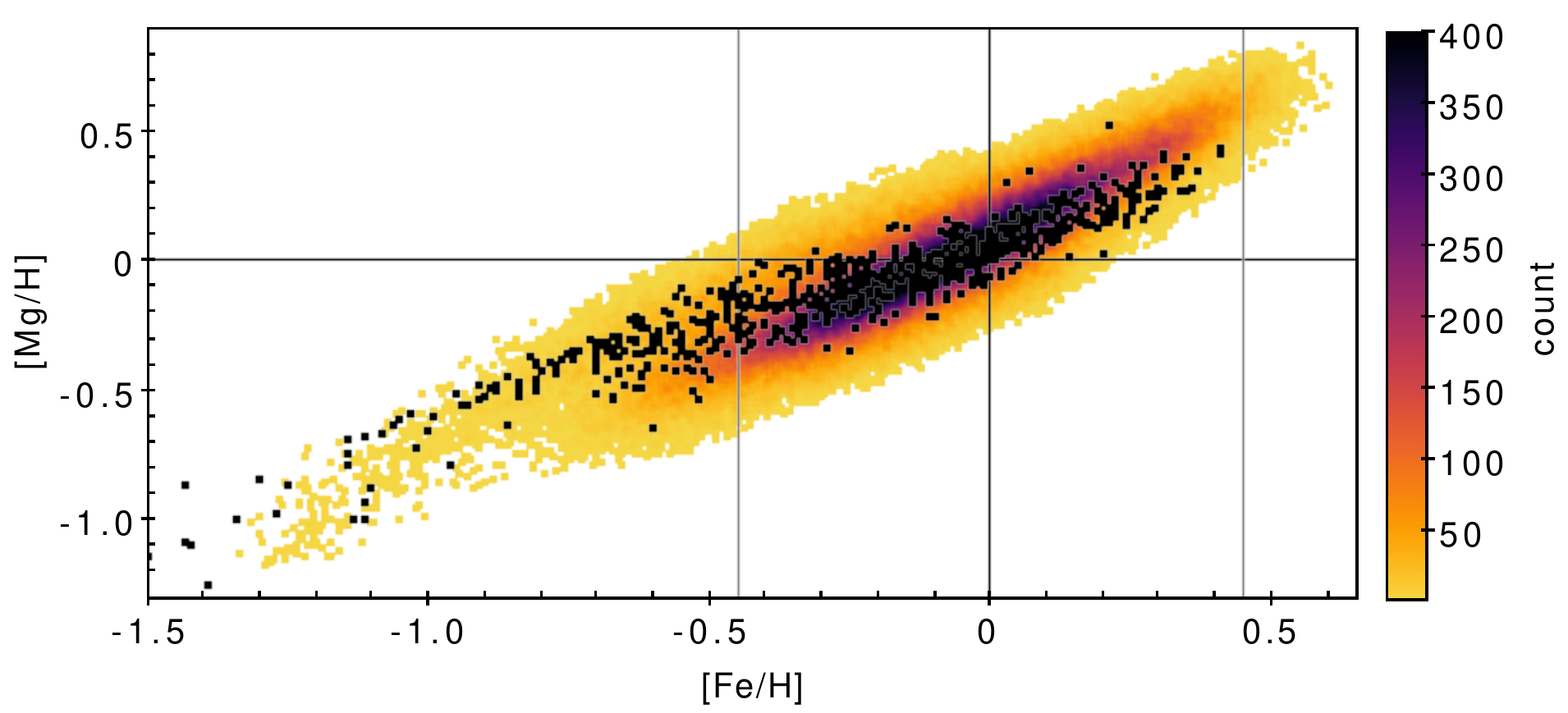}
 \includegraphics[scale=0.5]{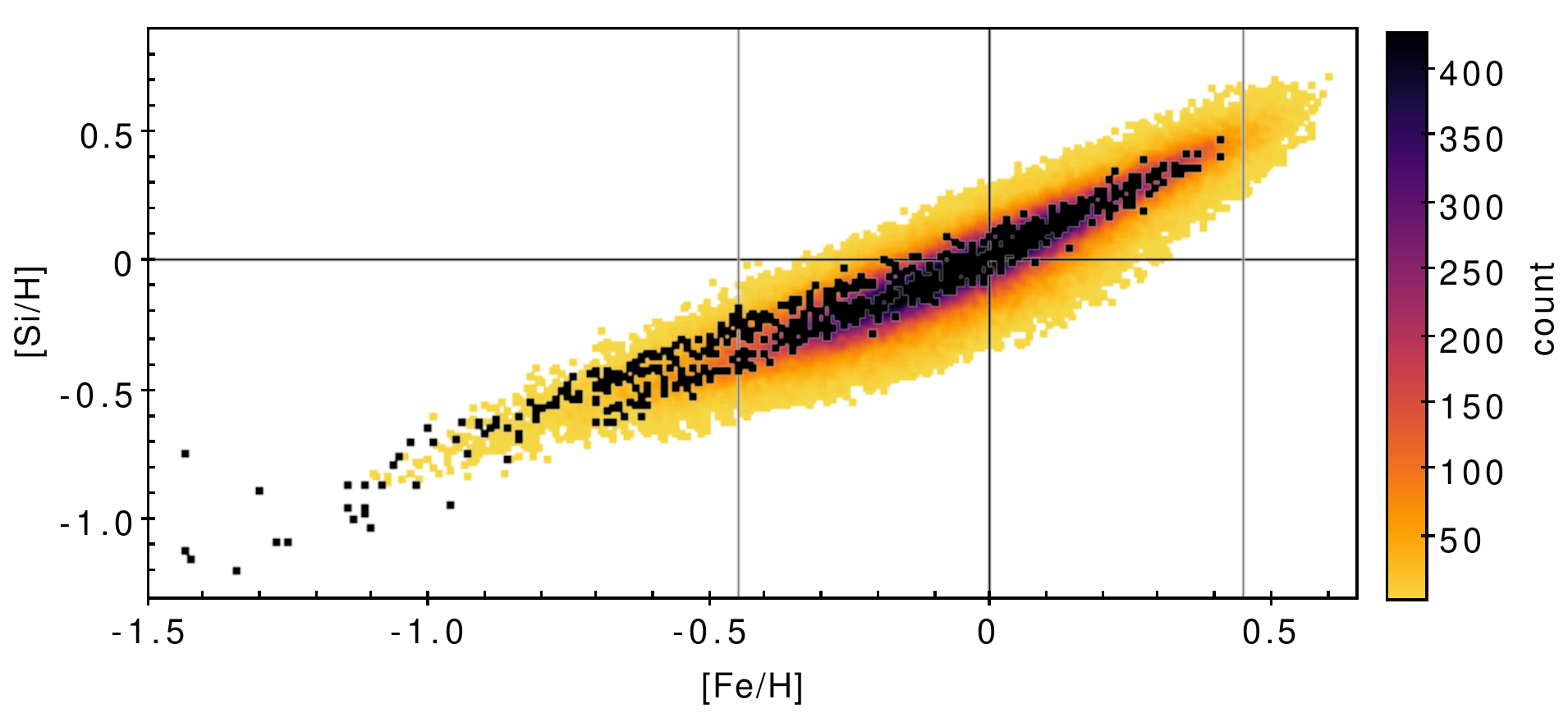}
 \caption{Carbon, oxygen, magnesium and silicon abundances for the stars in the GALAH sample \citep{2018MNRAS.478.4513B}. The black dots mark the data from the detailed survey of \citet{2014A&A...562A..71B}, which does not include carbon. This plot illustrates that the trends of the GALAH survey agree with previous studies and that there is some variations within the sample. The black vertical lines at [Fe/H]=-0.45 and 0.45 indicate the [Fe/H] span used in Fig.~\ref{fig:abundances}.
    \label{fig:stars}
   }
\end{figure}

We show in Fig.~\ref{fig:stars} the stellar abundances of carbon, oxygen, magnesium and silicon for the GALAH sample of stars used in our work. The overall trends agree with the previous study of \citet{2014A&A...562A..71B} and the values used for our calculations are shown in table~\ref{tab:abu} and plotted in Fig.~\ref{fig:abundances}.

\begin{table*}
\centering
\begin{tabular}{c|c|c|c|c|c|c|c|c|c}
\hline
[Fe/H] & $\sigma$[Fe/H] & [Si/H] & $\sigma$[Si/H] & [Mg/H] & $\sigma$[Mg/H] & [C/H] & $\sigma$[C/H] & [O/H] & $\sigma$[O/H]\\ \hline \hline
-0.4 & 0.02 & -0.32 & 0.06 & -0.26 & 0.08 & -0.29 & 0.05 & -0.19 & 0.08  \\  
-0.3 & 0.02 & -0.25 & 0.06 & -0.19 & 0.07 & -0.22 & 0.06 & -0.13 & 0.08  \\  
-0.2 & 0.02 & -0.17 & 0.06 & -0.10 & 0.07 & -0.14 & 0.06 & -0.08 & 0.08  \\  
-0.1 & 0.02 & -0.09 & 0.06 & -0.01 & 0.07 & -0.06 & 0.07 & -0.02 & 0.07  \\  
 0.0 & 0.03 &  0.00 & 0.06 &  0.08 & 0.07 &  0.03 & 0.07 &  0.03 & 0.07  \\ 
 0.1 & 0.02 &  0.10 & 0.06 &  0.19 & 0.07 &  0.11 & 0.07 &  0.08 & 0.07  \\ 
 0.2 & 0.02 &  0.20 & 0.06 &  0.30 & 0.07 &  0.20 & 0.07 &  0.13 & 0.07  \\ 
 0.3 & 0.02 &  0.32 & 0.05 &  0.42 & 0.06 &  0.28 & 0.06 &  0.19 & 0.07  \\ 
 0.4 & 0.03 &  0.41 & 0.05 &  0.53 & 0.06 &  0.36 & 0.06 &  0.23 & 0.07  \\  
\hline
\end{tabular}
\caption[Elemental abundances]{Stellar abundances as derived from the GALAH catalogue \citep{2018MNRAS.478.4513B}. This data is used to plot Fig.~\ref{fig:abundances}.}
\label{tab:abu}
\end{table*}

\section{Carbon contribution}
\label{ap:carbon}

The carbon fraction of minor bodies in the solar system is much higher than the carbon content within the Earth (e.g. \citealt{2015PNAS..112.8965B}), implying that the history of the chemical composition could change with orbital distance and planet formation models have to account for these effects. In our model, the water ice abundance is governed by the amount of free oxygen that does not form rock forming molecules or CO and CO$_2$ and not by orbital distance. Thus an elevated C/O ratio (Fig.~\ref{fig:CO}) reduces the amount of free oxygen that can form water ice. As a consequence the amount of carbon that forms carbon chain molecules without oxygen can influence the water ice content in our model.

\begin{figure}
 \centering
 \includegraphics[scale=0.7]{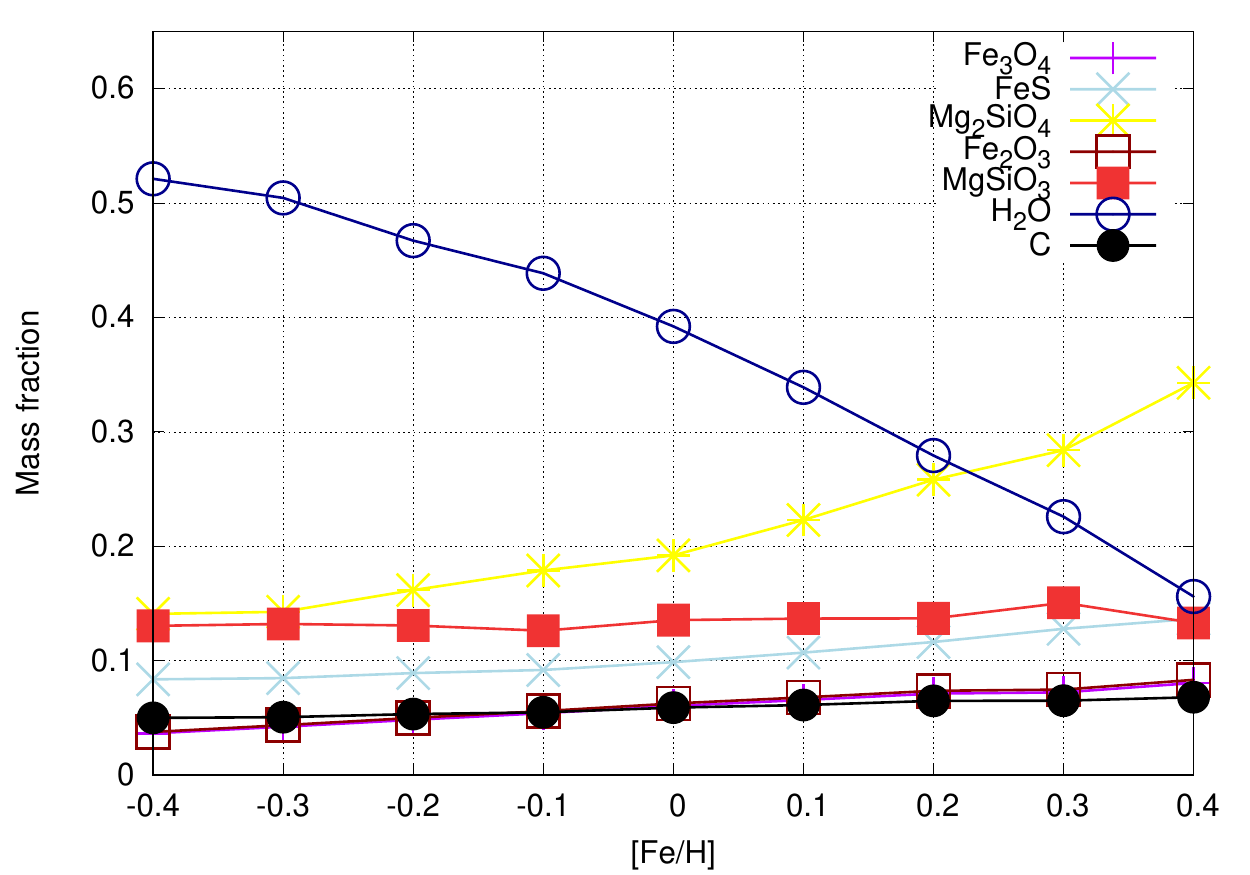}
 \caption{Molecular mass fractions of solid planetary building blocks formed exterior to the water ice line as function of [Fe/H] where 20\% of the C/H is assumed to be in carbon chains with an evaporation temperature of 500K and only 25\% of the C/H is assumed to be bound in CO. The trend for H$_2$O is similar to figure~\ref{fig:Moleculetrend}. However, the water ice abundance at large [Fe/H] is about a factor of 2-3 larger, while it stays roughly constant for low [Fe/H] compared to the nominal chemical model (Fig.~\ref{fig:Moleculetrend}).
    \label{fig:MoleculetrendlargeC}
   }
\end{figure}

In our nominal model (table~\ref{tab:species}) we included a fraction of 45\% of the C/H in organics (methane). We relax this approach here for testing purposes and reduce the CO content to 25\% and include a 20\% C/H fraction in pure carbon grains with an evaporation temperature of 500K (e.g. \citealt{2003A&A...410..611S}), bringing the total carbon content that forms non-oxygen bearing molecules to 65\% of the whole C/H, thus increasing the water content. We show these trends in Fig.~\ref{fig:MoleculetrendlargeC}.

Even with these assumptions, the general trend shown in Fig.~\ref{fig:Moleculetrend} holds and the water ice content decreases by nearly a factor of 3 from [Fe/H]=-0.4 to [Fe/H]=0.4. The difference to our nominal chemical model is larger for super-solar [Fe/H] compared to sub-solar [Fe/H]. This is caused by the low C/O ratio at low [Fe/H] (Fig.~\ref{fig:CO}), so that the effect of forming less CO compared to the total amount of oxygen that can form water is not significant. For high [Fe/H], the C/O ratio is much larger, resulting that the effect of forming less CO can actually increase the amount of oxygen that can from water. On the other hand, at high [Fe/H] also the rock forming species that bind oxygen like Fe, Si and Mg are more abundant than oxygen implying that the water ice ratio will decrease with increasing [Fe/H] independently of how much oxygen is bound in carbon to form CO or CO$_2$. The exact water content though can be estimated more accurately with more detailed chemical models, however, detailed models seem to indicate that the ration between CO to CO$_2$ to non-oxygen bearing carbon molecules is maximal equal, where in most cases the non-oxygen bearing carbon molecules are less abundant than CO and CO$_2$ \citep{2018A&A...613A..14E} which points to a water ice ratio as predicted by our nominal model (Fig.~\ref{fig:Moleculetrend}).

\section{Planet formation}
\label{ap:formation}

For our planet formation model we follow the core accretion scenario where planetary cores grow by pebble accretion \citep{2010A&A...520A..43O, 2010MNRAS.404..475J, 2012A&A...544A..32L} and migrate through type-I migration using the prescription of \citet{2011MNRAS.410..293P}. We use the disc model outlined in \citet{2015A&A...575A..28B}, where we either let the disc evolve in time or not to illustrate its influence on planetary composition similar to the principles outlined in \citet{2019A&A...624A.109B}. In addition we adopt two different $\alpha$ parameters for planetary migration to illustrate the effects of inward or outward migration close to the water ice line on the planetary composition as in \citet{2019A&A...624A.109B}.

The details of our pebble accretion model are described in \citet{2015A&A...582A.112B} and we just repeat here the necessities. We use a fixed pebble Stokesnumber $\tau_{\rm f}=0.1$ through the whole disc. The radial drift speed of the pebbles is given by
\begin{equation}
 v_{\rm r, peb} = 2 \frac{\tau_{\rm f}}{\tau_{\rm f} + 1} \eta v_{\rm K} \ ,
\end{equation}
where $\eta$ is the radial pressure gradient at the pebbles position and $v_{\rm K}$ is the Keplerian speed of the gas. We use a pebble flux of
\begin{equation}
 \dot{M}_{\rm peb} = 2 \times 10^{-4} \times 10^{[\rm Fe/H]} \exp(-t/10^6 {\rm yr} ) \frac{\rm M_{\rm E}}{\rm yr} \ ,
\end{equation}
where the component $10^{[\rm Fe/H]}$ describes a scaling with the host star metallicity. In addition we use an exponential decrease of the pebble flux in time $t$. With this pebble flux we can calculate the pebble surface density at the location of the planet $r_{\rm p}$ in the following way
\begin{equation}
 \Sigma_{\rm peb} = \frac{\dot{M}_{\rm peb}}{2 \pi r_{\rm p} v_{\rm r}} \ .
\end{equation}
The 2D Hill accretion rate onto the planetary core is then given as
\begin{equation}
 \dot{M}_{\rm core} = 2 r_{\rm H} v_{\rm H} \Sigma_{\rm peb} \ ,
\end{equation}
with $r_{\rm H}$ being the planetary Hill radius and $v_{\rm H} = \Omega_{\rm K} r_{\rm H}$ being the Hill speed. The planetary growth is stopped at the pebble isolation mass \citep{2014A&A...572A..35L, 2018arXiv180102341B, 2018A&A...615A.110A}, when the planet starts to carve a small gap in the protoplanetary disc which exerts a pressure bump exterior to it, stopping the inward flux of pebbles \citep{2006A&A...453.1129P} and thus pebble accretion. The slightly simplified pebble isolation mass \citep{2018arXiv180102341B} is given as
 \begin{equation}
  M_{\rm iso} = 25 \left(\frac{H/r}{0.05}\right)^{3} {\rm M}_{\rm E} \ .
 \end{equation}
As soon as the planet reaches its pebble isolation mass we stop pebble accretion, but at the same time we do not model the gas accretion component, because we are only interested in the composition of the planetary core.

The migration of the planet is determined by the formula of \citet{2011MNRAS.410..293P}. In this formula, the migration direction and speed depends crucially on the radial gradients of gas surface density and temperature/entropy. In addition outward migration by the entropy driven corotation torque can only happen, if the viscosity is large enough. In the disc model of \citet{2015A&A...575A..28B} this leads to a region of outward migration attached to the water ice line, if $\alpha_{\rm mig} > 0.001$, where the nominal value in this model is $\alpha_{\rm mig} = 0.0054$. We also test a variation with $\alpha_{\rm mig}=0.0001$ which results in only inward migration at the water ice line, changing the water content to smaller values, because the assembly of the planets can be finished interior to the water ice line \citep{2019A&A...624A.109B}. In section~\ref{sec:formation} we use $\alpha_{\rm mig} = 0.0054$, meaning that planets can migrate inwards and outwards. Here we show the results of different migration prescriptions and their influence on the water ice content of planets in a similar fashion as in \citet{2019A&A...624A.109B}.

\begin{figure}
 \centering
 \includegraphics[scale=0.7]{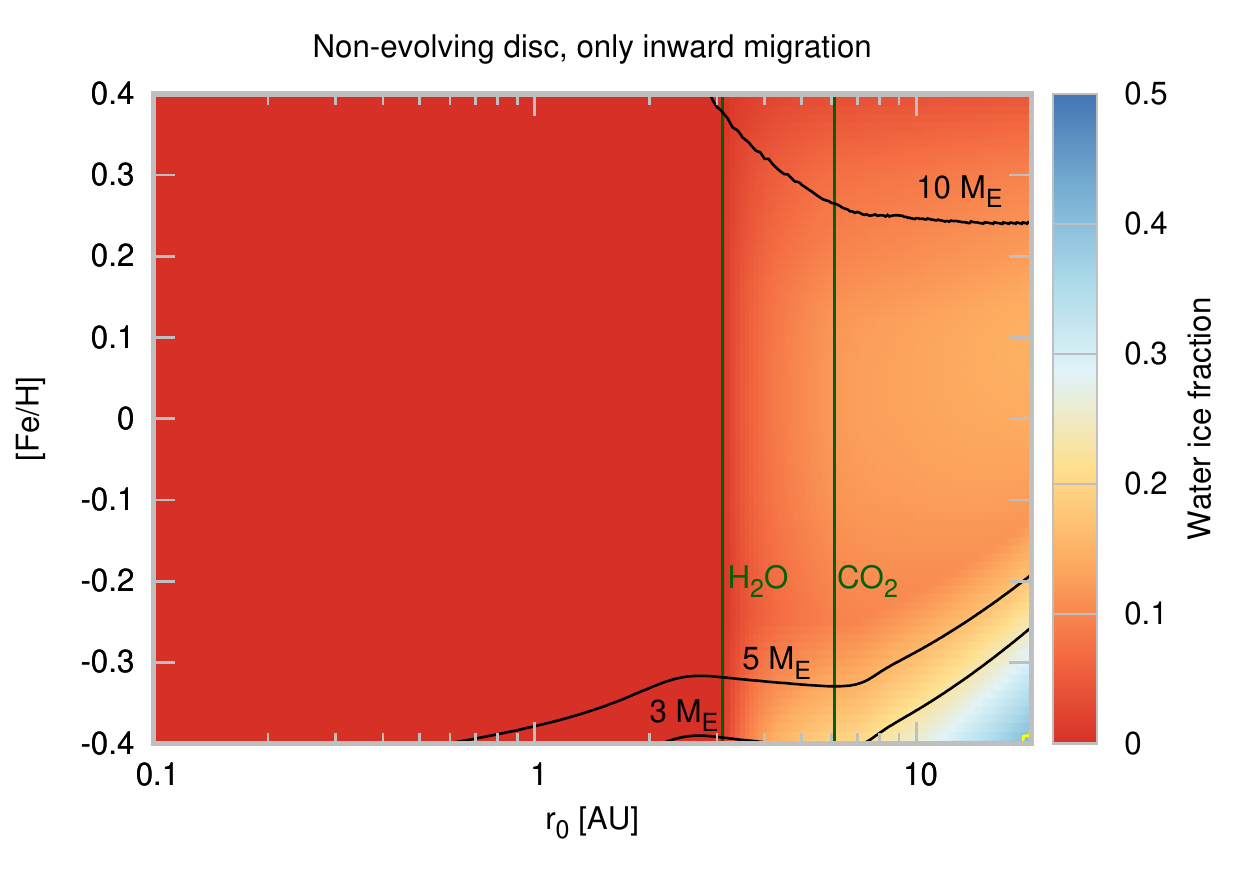}
 \includegraphics[scale=0.7]{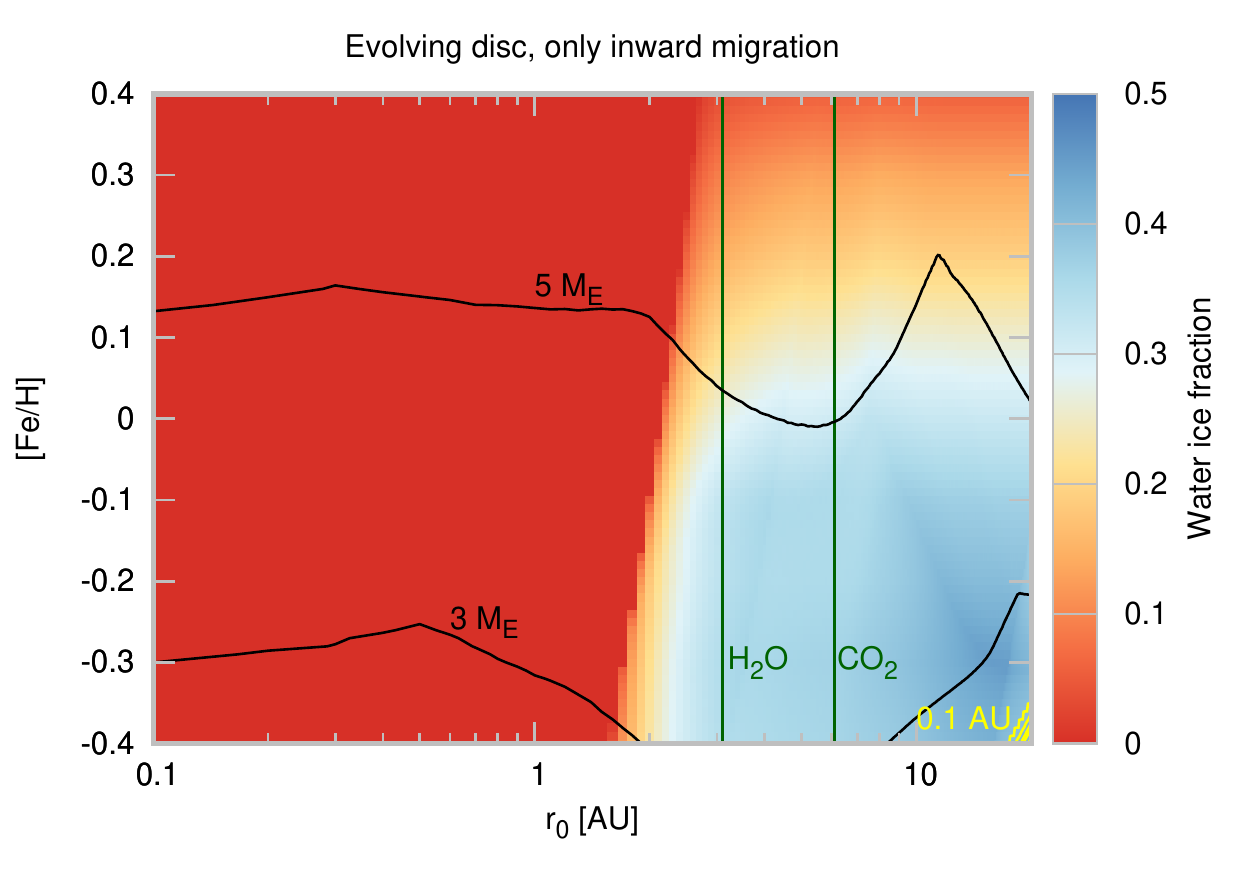}
 \includegraphics[scale=0.7]{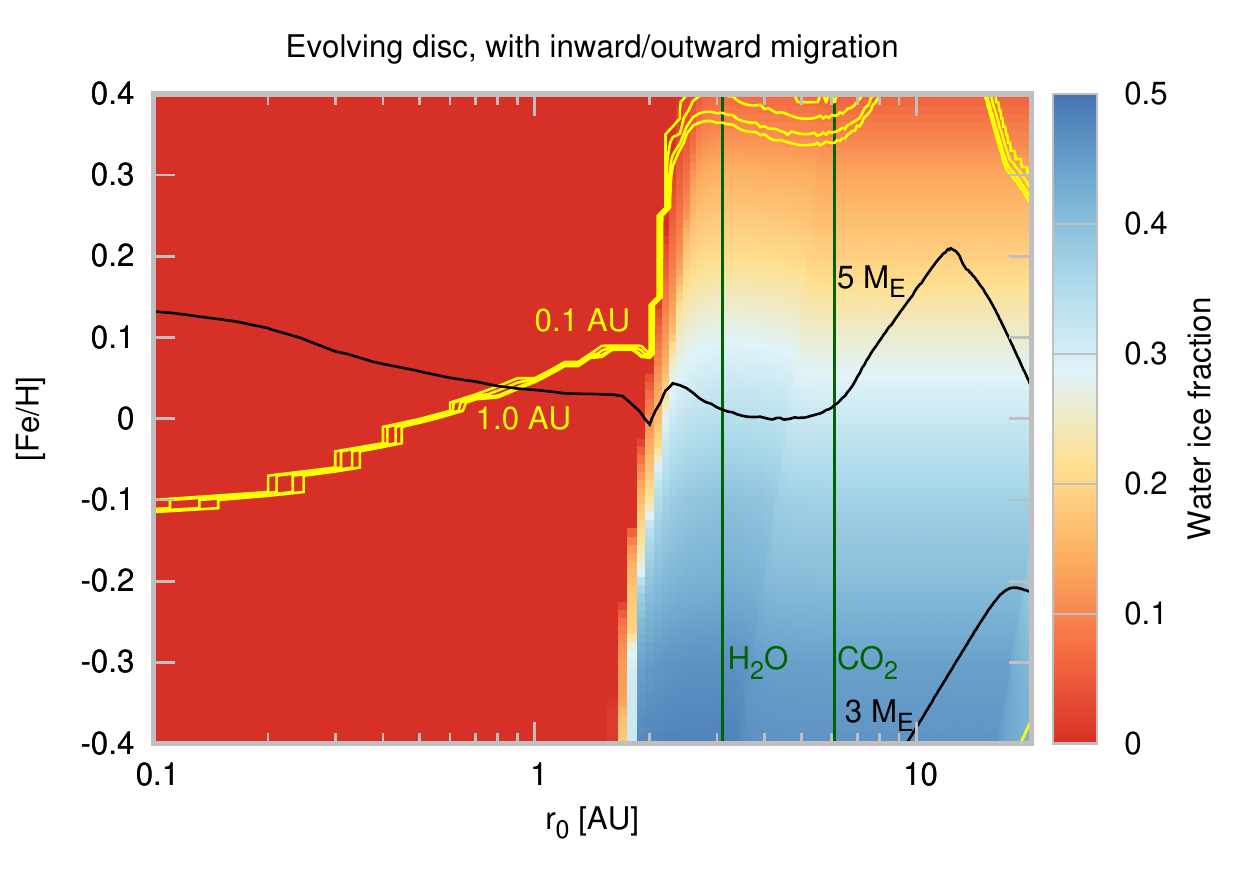}
 \caption{Final water ice content of planets formed with initial position $r_{\rm 0}$ and around stars with metallicity of [Fe/H] in either non-involving discs with only inward migration (top), evolving discs with only inward migration (middle) and evolving discs with inward and outward migration (bottom). Planets are injected at $t_0$=0.5 Myr. The black and yellow lines mark the final orbital position (0.1, 0.3, 0.5 and 1.0 AU) and masses of the planets. In the simulations presented in the top two panels, nearly all planets migrate to the inner edge at 0.1 AU. The green vertical lines depict the H$_2$O and CO$_2$ ice lines at the beginning of the simulations. H$_2$O and CO$_2$ are in icy form exterior to these ice lines. Due to the disc's evolution in time (middle and bottom), the water ice line moves inwards, so that the water ice line can sweep over slow growing planets initially interior to the water ice line allowing them to have a water ice contribution.
    \label{fig:H2Ocontent}
   }
\end{figure}

We place planetary embryos of initially 0.01 Earth masses at different initial positions $r_{\rm 0}$ in discs that are initially already 0.5 Myr old (in order to account for the formation of the planetary embryo). We then vary the [Fe/H] of the host star and calculate the water ice fraction of the formed planets via our nominal chemical model. In Fig.~\ref{fig:H2Ocontent} we show the water ice content of formed planets in our model for three different assumptions, (i) a non-evolving disc with only inward planetary migration modelled with $\alpha_{\rm mig}=0.0001$ (top), (ii) an evolving disc with only inward planetary migration (middle) and (iii) an evolving discs where planets can migrate inwards and/or outwards with $\alpha_{\rm mig}=0.0054$ (bottom). The final planetary masses are between 3 and 10 Earth masses in all models and if planetary migration is only inwards, planets reach the inner edge at 0.1 AU, while in the scenario with outward migration, planets are trapped at around 1.8AU, which corresponds to the outer edge of the region of outward migration in that disc model \citep{2015A&A...575A..28B}.

For the non-evolving disc with only inward migration, the results are as expected. Planets starting to form exterior to the water ice line initially accrete water ice and then migrate into the inner disc where they finish their assembly and thus feature a low water ice content. All planets in this case reach 0.1 AU. Planets forming exterior to the CO$_2$ ice line feature a higher water ice content, because they can finish most of their growth exterior to the water ice line and thus accrete water ice. This is also reflected by the water ice gradient of the planets with increasing initial planetary position, meaning that the farther the planetary embryo is placed exterior to the water ice line, the higher its water ice content. This result is consistent with \citet{2019A&A...624A.109B} derived for solar metallicity. In addition with increasing [Fe/H] the water ice content decreases, as shown in the main paper.

In case the disc evolves, the water ice line moves inward in time due to the decrease in viscous heating and even planetary embryos that are initially placed interior to the water ice line can accrete water ice, because the evolution of the water ice line is initially faster than planet migration \citep{2019A&A...624A.109B}. If planetary accretion is faster, planets grow quicker and they can migrate inwards faster, avoiding the sweep of the inward moving water ice line. This is for example the case at high [Fe/H], where the pebble flux is larger and thus accretion is faster and only planets initially slightly interior to the water ice line can accrete some water ice. The planets in this scenario all migrate to the inner edge of the disc at 0.1 AU.

In the case of an evolving disc where planets can migrate inward and outwards, the water ice content of the fully grown planets is highest of these three models. This is caused by the fact that planets in this model migrate outwards exterior to the water ice line where they then finish their assembly and thus accrete all their mass with a water ice component \citep{2019A&A...624A.109B}. In the end, planets either migrate to the inner edge of the disc at 0.1 AU or stay trapped in the region of outward migration exterior to 1 AU until disc dissipation. The water ice component of the planets decreases slightly for planetary seeds forming exterior to a few AU, because the disc becomes so cold that the planets also accrete a fraction of CO$_2$ ice. This decreases the relative water ice component of the accreted material and results in a lower water ice fraction of the formed planets.

In any case, all these simulations clearly show that the water ice content of the formed planets is a mixture and always in between the states of zero water ice (planets formed completely interior to the water ice line) or up to the maximal water ice content allowed in the chemical model (planets that form completely exterior to the water ice line). The planet formation simulations also clearly demonstrate the water ice gradient from low [Fe/H] to high [Fe/H], as outlined in the main part of the paper and section~\ref{sec:formation}.

\bibliographystyle{aa}
\bibliography{Stellar}
\end{document}